\documentclass[sn-mathphys,Numbered]{sn-jnl}

\usepackage{graphicx}%
\usepackage{multirow}%
\usepackage{amsmath,amssymb,amsfonts}%
\usepackage{amsthm}%
\usepackage{mathrsfs}%
\usepackage[title]{appendix}%
\usepackage{xcolor}%
\usepackage{textcomp}%
\usepackage{manyfoot}%
\usepackage{booktabs}%
\usepackage{algorithm}%
\usepackage{algorithmicx}%
\usepackage{algpseudocode}%
\usepackage{listings}%

\theoremstyle{thmstyleone}%
\theoremstyle{thmstyletwo}%

\theoremstyle{thmstylethree}%

\begin{document}

\title[Data-Centric Digital Agriculture: A Perspective]{Data-Centric Digital Agriculture: A Perspective}


\author*[1,2]{\fnm{Ribana} \sur{Roscher}}\email{r.roscher@fz-juelich.de}

\author[3]{\fnm{Lukas} \sur{Roth}}\email{lukas.roth@usys.ethz.ch }

\author[4]{\fnm{Cyrill} \sur{Stachniss}}\email{cyrill.stachniss@igg.uni-bonn.de}

\author[3]{\fnm{Achim} \sur{Walter}}\email{achim.walter@usys.ethz.ch}

\affil*[1]{\orgdiv{Data Science for Crop Systems}, \orgname{Forschungszentrum Jülich GmbH}, \orgaddress{\street{Wilhelm-Johnen-Straße}, \city{Jülich}, \postcode{52428}, \country{Germany}}}

\affil[2]{\orgdiv{Remote Sensing Group}, \orgname{University of Bonn}, \orgaddress{\street{Niebuhrstr. 1a}, \city{Bonn}, \postcode{53113}, \country{Germany}}}

\affil[3]{\orgdiv{Institute of Agricultural Sciences}, \orgname{ETH Zurich}, \orgaddress{\street{Universitätstrasse 2}, \city{Zürich}, \postcode{8092}, \country{Switzerland}}}

\affil[4]{\orgdiv{Photogrammetry \& Robotics Lab}, \orgname{University of Bonn}, \orgaddress{\street{Nussallee 15}, \city{Bonn}, \postcode{53115}, \state{NRW}, \country{Germany}}}


\abstract{In response to the increasing global demand for food, feed, fiber, and fuel, digital agriculture is rapidly evolving to meet these demands while reducing environmental impact. This evolution involves incorporating data science, machine learning, sensor technologies, robotics, and new management strategies to establish a more sustainable agricultural framework.
So far, machine learning research in digital agriculture has predominantly focused on model-centric approaches, focusing on model design and evaluation. These efforts aim to optimize model accuracy and efficiency, often treating data as a static benchmark. Despite the availability of agricultural data and methodological advancements, a saturation point has been reached, with many established machine learning methods achieving comparable levels of accuracy and facing similar limitations.
To fully realize the potential of digital agriculture, it is crucial to have a comprehensive understanding of the role of data in the field and to adopt data-centric machine learning. This involves developing strategies to acquire and curate valuable data and implementing effective learning and evaluation strategies that utilize the intrinsic value of data.
This approach has the potential to create accurate, generalizable, and adaptable machine learning methods that effectively and sustainably address agricultural tasks such as yield prediction, weed detection, and early disease identification.}




\maketitle

\section{Introduction}
Agriculture is at the forefront of the world's most critical challenges, including securing food, feed, fiber, and fuel, as well as addressing climate change. Therefore, it plays a crucial role in achieving the United Nations' Sustainable Development Goals, specifically SDG 2 (no hunger) and SDG 13 (climate action). It is responsible for 15\%-25\% of global greenhouse gas emissions~\cite{vermeulen2012climate} and is susceptible to the impacts of droughts and floods \cite{Devot2023}. The decline in biodiversity and ecosystem services further highlights the need for improved field management practices, fertilizers, and pesticides \cite{ortiz2021review}. 
Traditional agriculture, which relies on farmers' intuitions and experiential decision-making, is increasingly challenged by consumers' and citizens' expectations of sustainable agriculture \cite{Walter2017}. 
Modern agriculture, particularly digital agriculture, has emerged as a crucial approach to address these challenges. Digital agriculture utilizes various technologies such as sensors and robotics and primarily relies on data-driven methods to optimize agricultural practices \cite{trendov2019digital}. The main aim is to inform decision-making regarding optimal and sustainable resource allocation, appropriate amounts, places, and timings of measures. Farmers, researchers, and policymakers can increasingly access diverse sources of observational and simulation-based data, such as soil nitrogen availability, rainfall forecasts, crop development and yield potential, disease risks, and weed or pest densities. Insights derived from this data can be translated into actionable responses, leading to enhanced sustainability in agricultural management practices \cite{basso2020digital}.
The transformative potential of digital agriculture is driven by the synergistic integration of digital technologies, along with access to data through the use of new data sources, efficient and targeted data collection, and the ability to optimize data quality and usage. This perspective emphasizes the often-overlooked importance of data in digital agriculture and introduces the concept of \emph{data-centric digital agriculture} with a specific focus on crop production. It explores promising developments in data science, sensor technologies, and robotics that are crucial to realizing the full potential of data.

\section{From Early Precision Agriculture via Farm Management Systems to Agricultural Robotics}
Over several decades, agriculture has transformed significantly, with data and technology becoming central elements \cite{kamilaris2017review,coble2018big,shamshiri2018,carletto2021agricultural}. These innovations have reshaped farming management and the field from traditional to modern agriculture. In this section, we will elaborate on this evolution, from the beginnings of precision agriculture through the rise of digital agriculture to agricultural robotics. We will discuss how the combination of data, technological advancements, and innovative farming practices have collectively changed the agricultural landscape, leading to more informed decision-making and raising exciting questions about farming's future.

The nature and variety of data sources in agriculture have evolved and expanded significantly (as detailed in Box 1). Early agricultural data was confined to manual records and observational notes, often documented in books and logbooks. While valuable, these records lacked the depth and scope of today's data sources.
Modern agriculture has seen a proliferation of data types, including experimental data from controlled trials and research adhering to strict scientific protocols. Moreover, non-experimental data collected from real-world farming scenarios have become invaluable due to their practical applicability.
Simultaneously, the integration of various sensors and technologies into agriculture has increased. These technologies range from soil and weather sensors to drone-based remote sensing and satellite imagery, enhancing the granularity and real-time aspect of data acquisition.
The trend towards a richer and more diverse range of data sources is expected to continue. In combination with digital technologies, it promises to deepen our understanding of agricultural systems. This evolution not only reshapes the data landscape in agriculture but also highlights the potential for more precise, sustainable, and efficient farming practices.

\begin{figure}[t]
    \centering
    \includegraphics[width=1\textwidth]{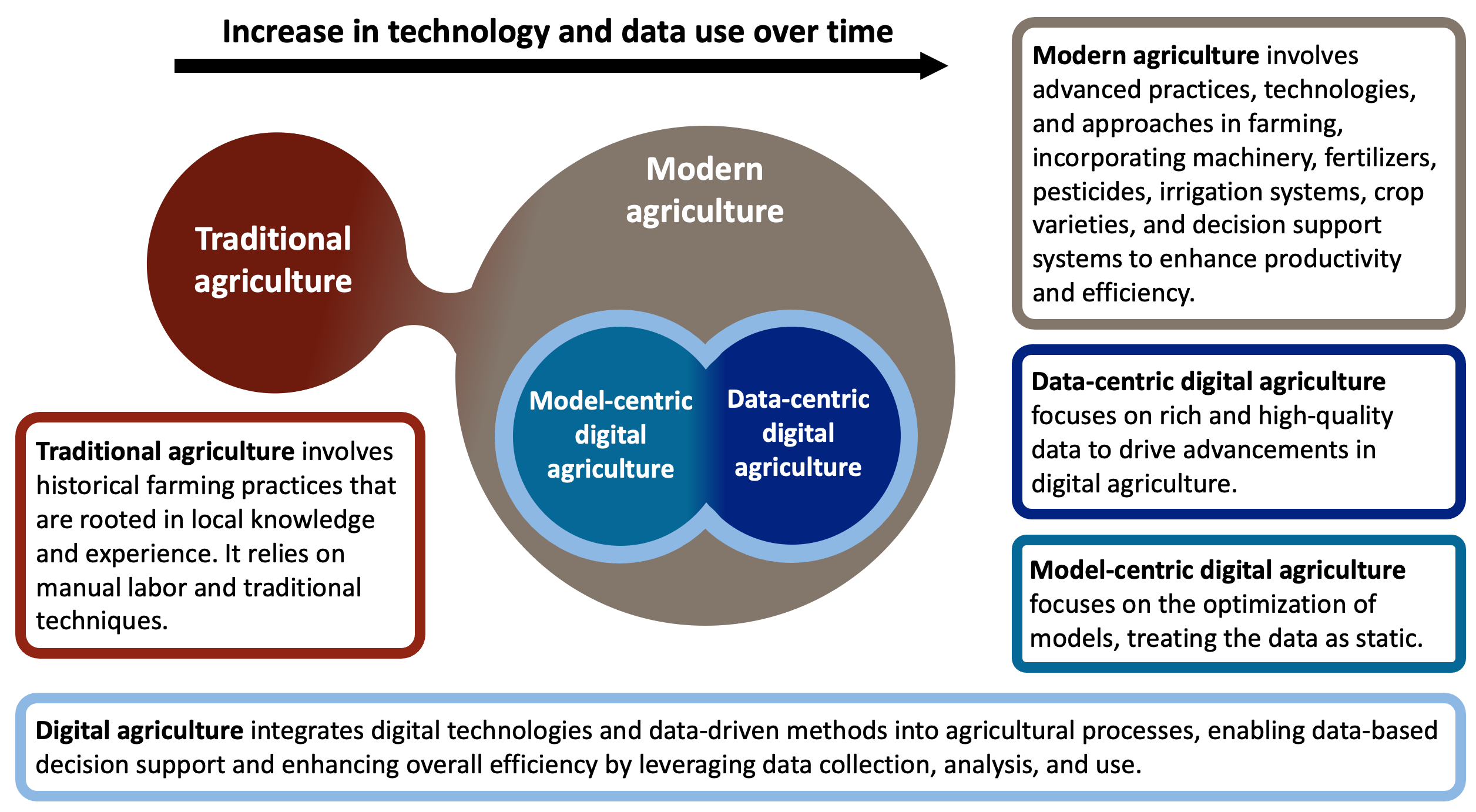}
    \caption{In this Venn diagram, we visually represent the evolution of different agricultural paradigms over time as technology usage and available data increase. In the past, traditional agriculture was dominant, but currently, the focus has shifted to digital agriculture. In the future, we expect data-centric digital agriculture to become increasingly important. It is worth noting that different paradigms can coexist, and a blending of different practices takes place as they evolve. This underlines the dynamic and interconnected nature of modern agriculture.}
    \label{fig:terminology}
\end{figure}

\begin{figure}[t]
    \centering
    \includegraphics[height=3.3cm]{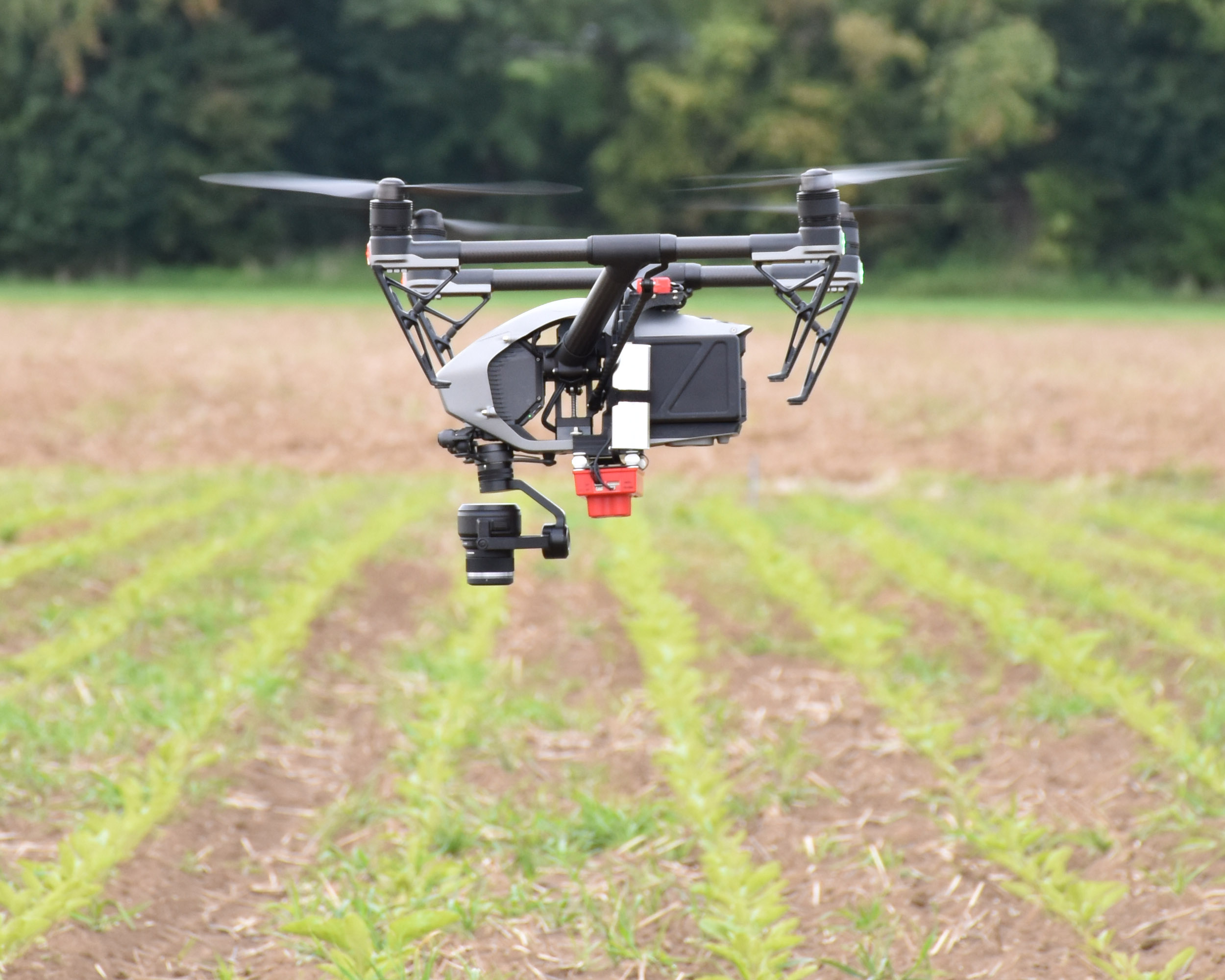}\,
    \includegraphics[height=3.3cm]{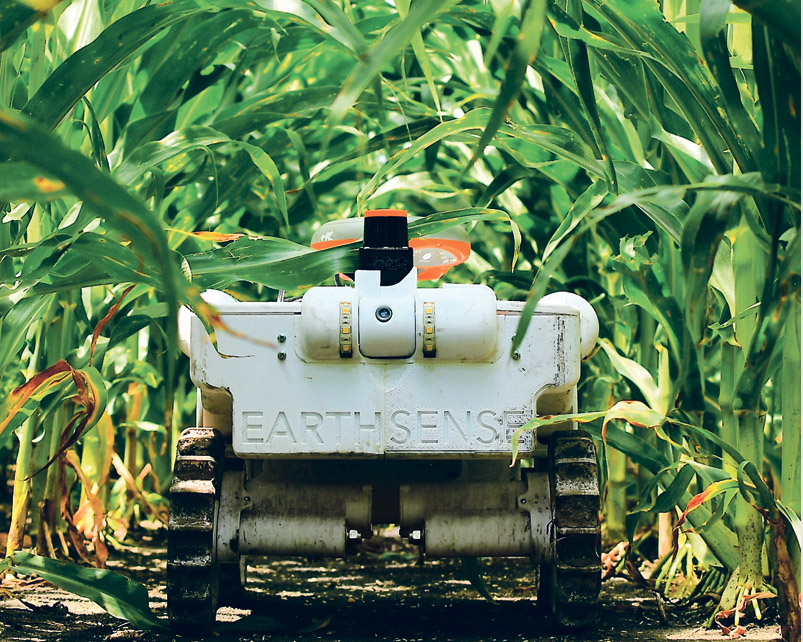}\,
    \includegraphics[height=3.3cm]{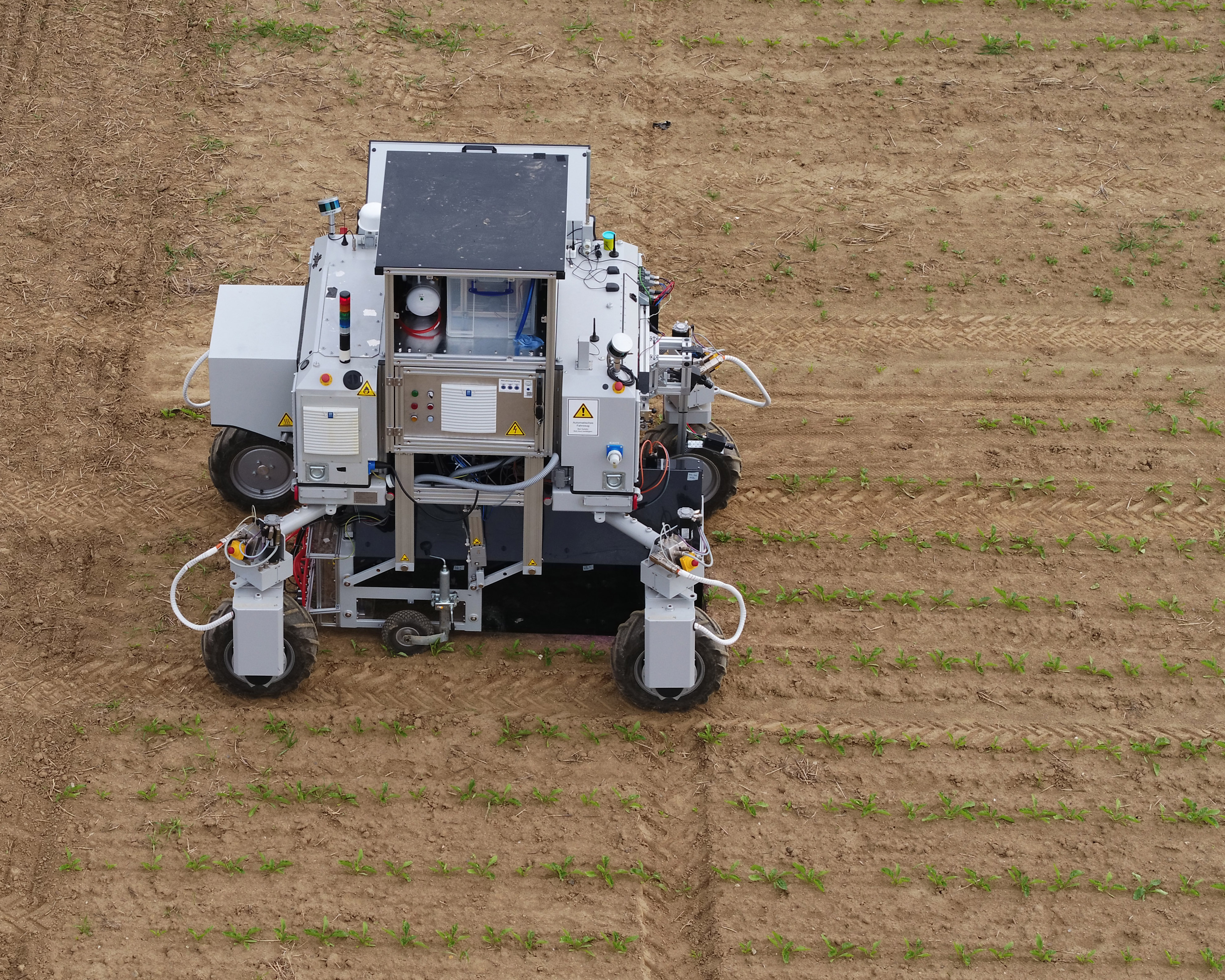}\\[1mm]
    \includegraphics[height=3.3cm]{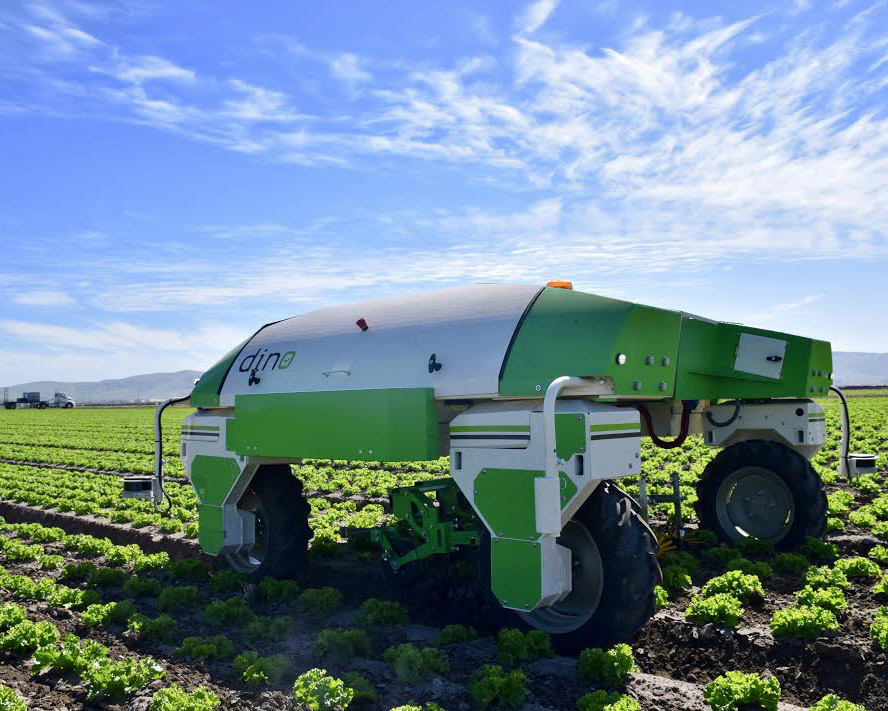}\,
    \includegraphics[height=3.3cm]{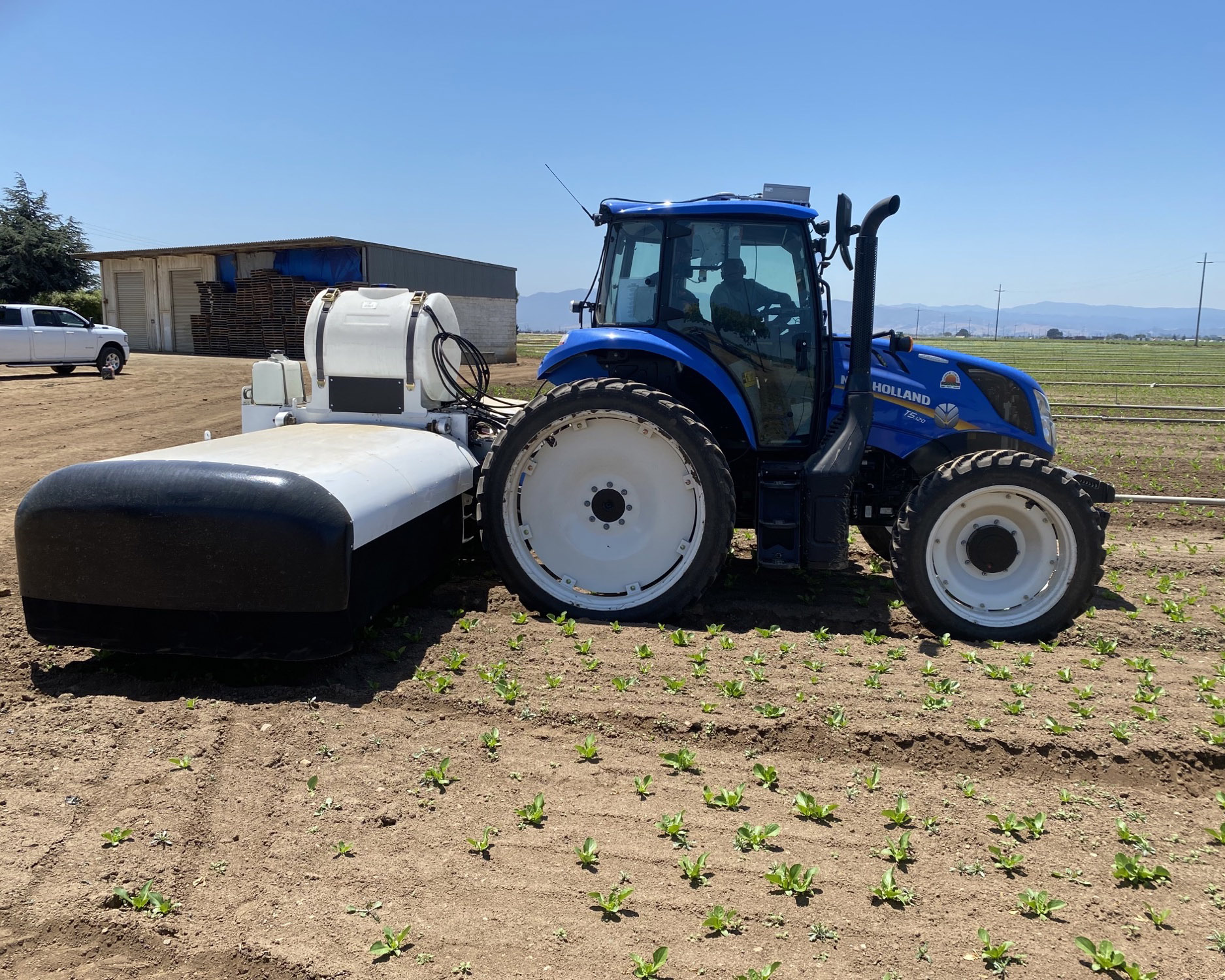}\,
    \includegraphics[height=3.3cm]{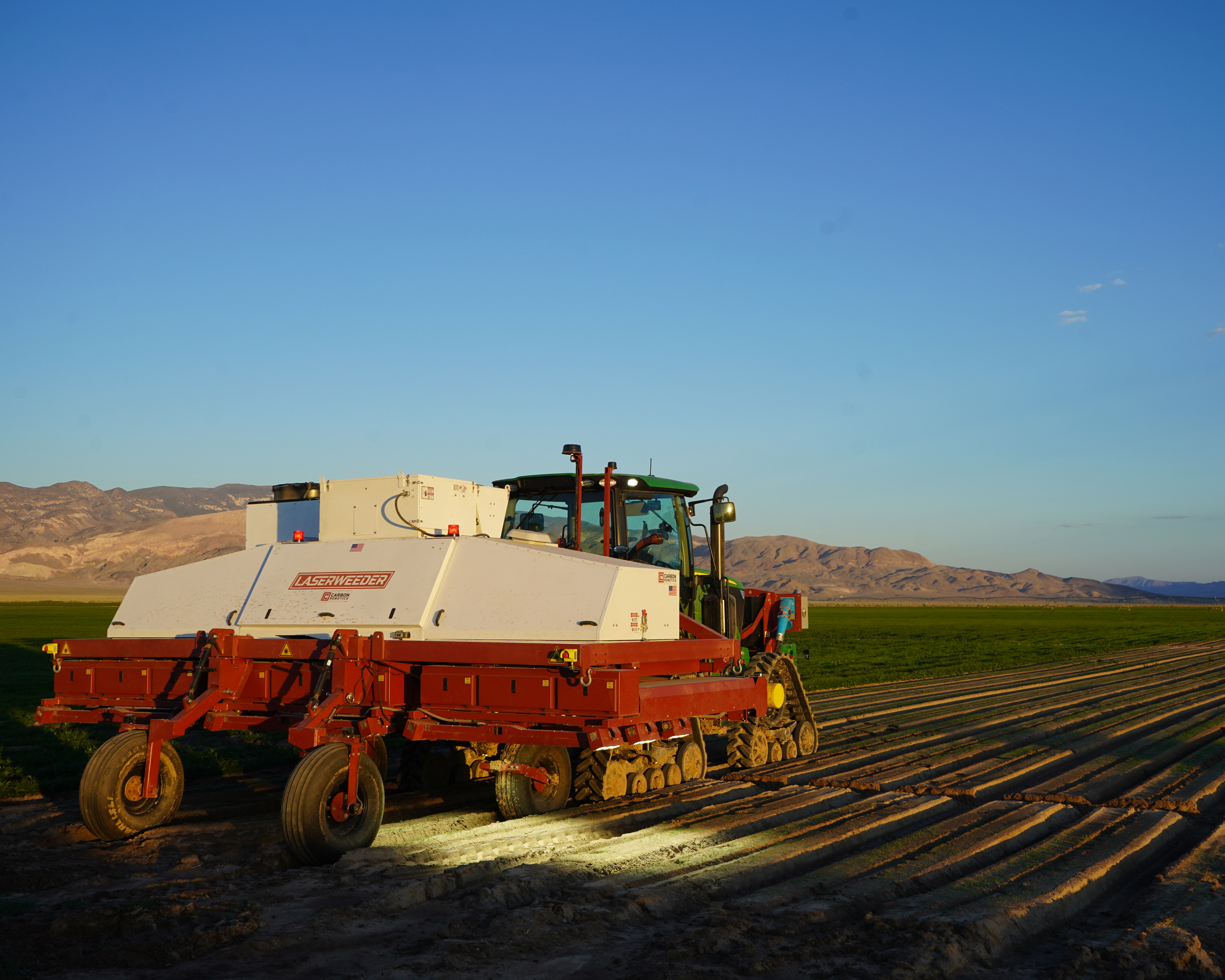}
    \caption{Unmanned aerial vehicles (UAVs), robots, and smart agricultural machines integrate monitoring, phenotyping, spot-spraying, and chemical-free weeding. Image courtesy: PhenoInspect GmbH, EarthSense, Flourish Project, Naio, Verdant, Carbon Robotics.}
    \label{fig:robots}
\end{figure}

Alongside this data evolution, the increasing availability of advanced digital technologies and data sources, coupled with the potential for improved data usage, has led to new paradigms, as outlined in Fig. \ref{fig:terminology}.
Traditionally, farmers based their field management decisions on qualitative observations at the field-level. 
In the mid-80s, precision agriculture applications emerged and gained traction later in the 90s \cite{gebbers2010,lowenberg2015precision}.
Two developments were fundamental prerequisites for their establishment: (1) The ability to resolve in-field heterogeneity in growth conditions, such as plant-available nutrients in the soil, and growth status, such as plant biomass, and (2) the ability to apply spatially resolved treatments to these heterogeneities, for example, variable rate application of nutrients.
The broader field of digital agriculture has emerged more recently due to advances in digital technologies such as data science, machine learning, and robotics \cite{trendov2019digital,basso2020digital}. While precision agriculture can be considered a subset or component of digital agriculture, precision agriculture was the original approach that laid the foundation for digital agriculture. It paved the way for integrating digital technologies and data-driven practices into agriculture and enabled the development of more comprehensive and advanced digital agriculture solutions (see Fig~\ref{fig:robots}). 

In the field of digital agriculture, phenotyping has become a significant focus area \cite{li2020review,yang2020crop,saric2022applications}. Phenotyping in agriculture involves the detailed measurement and analysis of observable plant traits. These traits encompass a wide range of characteristics, including growth patterns, stress responses, and crop yield predictions.
In the past, phenotyping relied on qualitative assessments of these traits. However, advancements in sensor technology and data science strategies, particularly in imaging and remote sensing, have transformed phenotyping into a data-rich and quantitative field. Through the use of advanced sensor technologies, high-throughput phenotyping allows for precise and detailed measurements of plant traits. This brings a new dimension to digital agriculture by providing real-time, data-driven insights into plant health and performance.

The agricultural transformation extends beyond just data and technology. Advancements in digital technologies significantly impact farm management and crop cultivation methods.
Practical solutions such as crop diversification, along with new field arrangements and smaller individual field sizes, can be important steps to address the multiple challenges of agriculture. Smaller, homogeneous crop patches can benefit biodiversity, pest control, and soil fertility \cite{hernandez2022model}. Experiments with 3-m-wide strip cropping systems \cite{juventia2022spatio}, flowering strips \cite{tschumi2015high}, hedgerows, contour plowing, and other biodiversity-friendly land-use practices \cite{tscharntke2021beyond} have shown increased ecosystem services. However, diversified field arrangements require an appropriately high data resolution, which could be achieved with high-quality field sensor data and remote and proximal sensing-based devices. Moreover, mechanistic and machine learning-based agroecosystem models simulating crop growth, development, and soil processes in response to environmental conditions and management practices will become more crucial \cite{rotter2015use,muller2019plant,drees2021temporal,miranda2022controlled}. 

\begin{figure}[t]
    \centering
    \includegraphics[height=3.3cm]{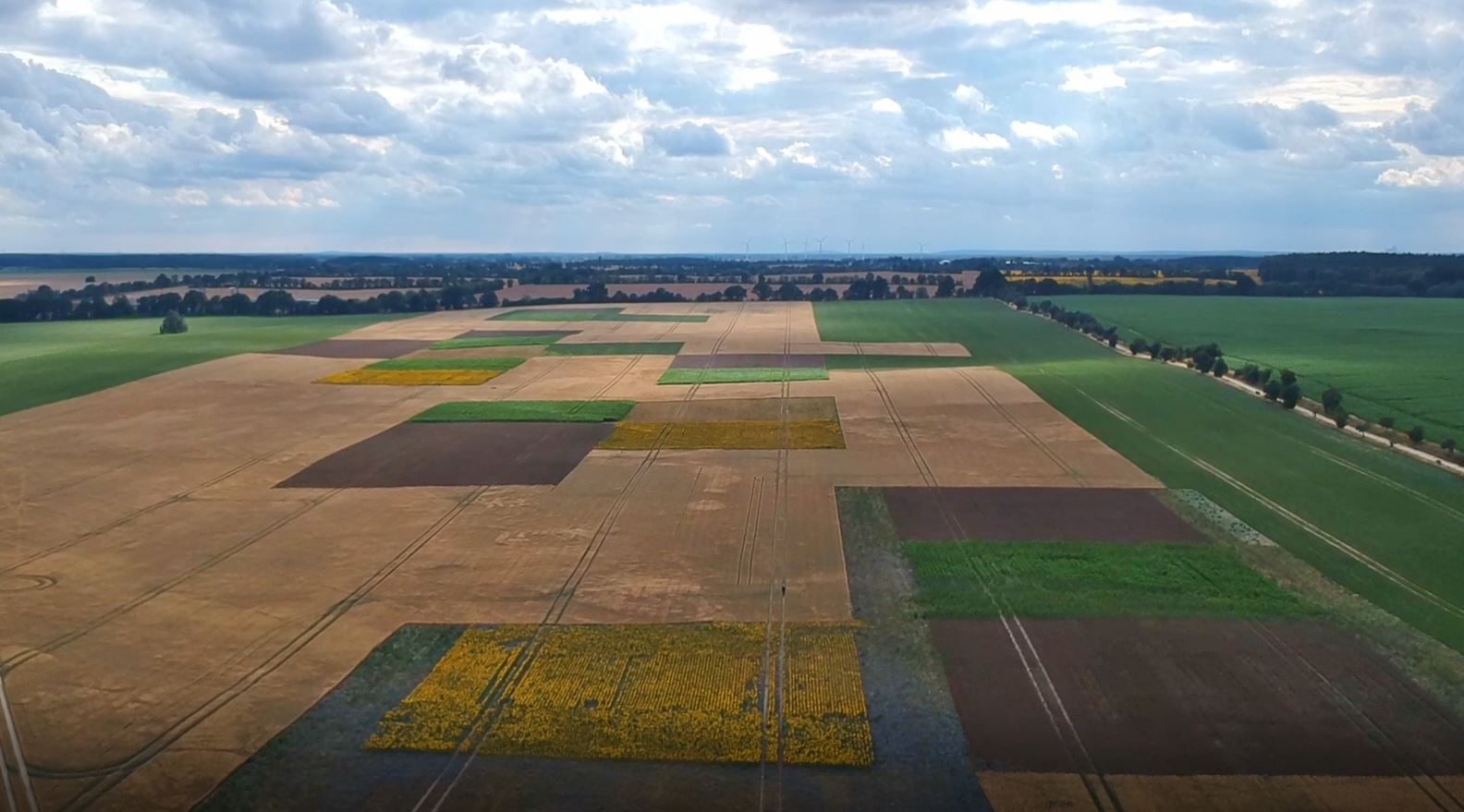},
    \includegraphics[height=3.3cm]{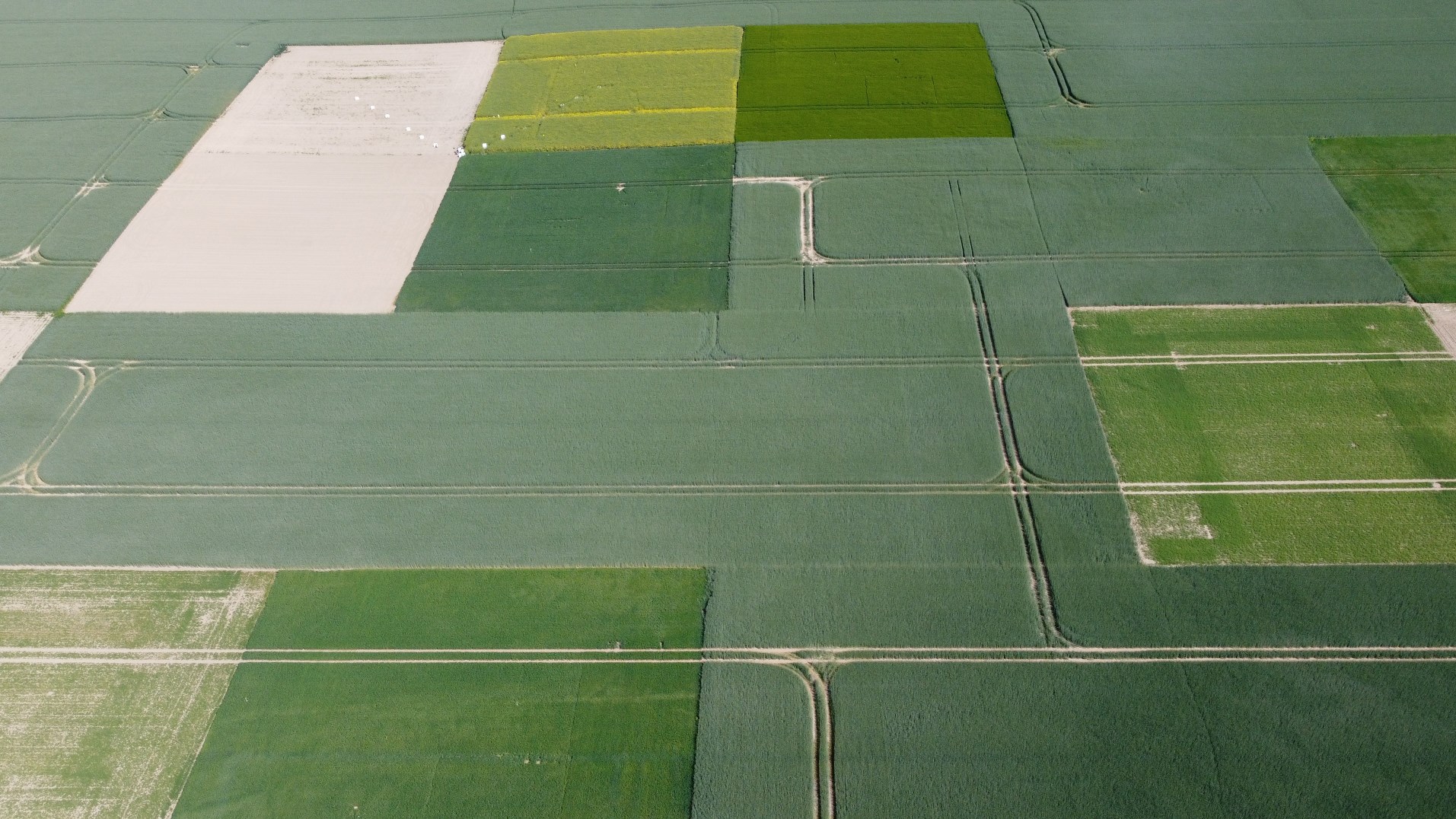}
    \caption{Innovative cropping systems such as patch cropping focus on a high plant diversity in small areas, fostering that crops can grow where they find optimal conditions for growth in the soil. In this way, chemical synthetic pesticides and fertilizers can be reduced. The image shows the patchCROP experiment at Leibniz Centre for Agricultural Landscape Research (ZALF) Müncheberg, with permanent tracks and a GPS-controlled control system that enable targeted and precise trials. Image courtesy: Hendrik Schneider, Marten Schmidt/ZALF}
    \label{fig:patch}
\end{figure}

If fields are managed in patches according to their overall natural boundary conditions (soil properties, sun exposure, and other microclimatic factors), choosing the ideal cultivar from the available spectrum becomes even more critical (see Fig. \ref{fig:patch}). Similarly, segmenting an entire field or farm into appropriate patches for crop rotation \cite{donat2022} would be a significant challenge. These challenges can be effectively addressed with the capabilities of various sensing technologies and data processing and analysis. For example, multi-year yield assessments can be conducted to determine the appropriate patches for the field. These assessments can be done using combined harvester data or retrospectively using satellite-based yield estimations \cite{perich2023pixel}. Managing fields with irregular sizes is only feasible if tractors and agricultural machinery are GPS-controlled or even autonomous to reduce labor. Coupling these management procedures with controlled-traffic farming, where tractors repeatedly use the same lanes in the field, can minimize the effects of soil compaction as much as possible \cite{donat2022}.

Despite four decades of development in digital agriculture, it remains confined to a few use cases, raising questions about its broader adoption \cite{kamilaris2017review}. One observation we make is that although there are numerous machine learning methods and statistical tools available, the lack of widely applicable models is hindering the progress of digital agriculture. Currently, data-driven autonomous solutions are primarily employed in tasks related to established crops and varieties that are typically managed using large machinery~\cite{finger2019precision}. However, to reduce environmental impact and foster stronger connections between producers and consumers, there's growing interest in envisioning a future with a small fleet of lightweight robots and intelligent assistants supporting farmers. Such solutions have the potential to efficiently manage diverse crops on small plots, potentially revolutionizing sustainable cultivation practices.

Another critical observation we make is that the development of field management techniques and machinery in agricultural practices has not kept pace with the advancements in sensing technologies. A prime example of this is mineral nitrogen (N) fertilizer, one of the most crucial farming inputs, as elaborated in Box 2.
Commercially available sensing products now possess precision that surpasses the capabilities of most conventional fertilizer spreaders, leading to discrepancies between sensing and application, which may be exacerbated by the shift from proxy measurements like vegetation indices to single plant and organ detection techniques \cite{weyler2021joint,marks2022precise,shaikh2022towards}.
These advancements hold promise for tasks such as weed control and pest management. However, despite research demonstrating the feasibility of site-specific treatments, commercial systems are limited for large-scale operations. Consequently, the recent emphasis in digital agriculture has been on increasing the efficiency of large-scale farming machinery~\cite{lottes2018,pretto2021}. 
This focus on 'economies of scale' has resulted in larger farms being managed by the same workforce as before~\cite{finger2019precision}. Yet, it is essential to recognize that ecological improvements cannot be solely achieved by reducing human inputs; conserving other natural resources is equally critical.

Based on these observations, we state that the future of digital agriculture depends on several critical factors, but above all, the trade-off between costs and benefits must improve. One challenging yet readily attainable goal we can strive for is to develop agricultural models that are accurate and provide a high generalization power and adaptability to be applicable to various agricultural scenarios and tasks. We argue that achieving this goal involves improving data creation, selection, curation, and utilization, potentially through different machine learning paradigms. Further, it is essential to understand better which data holds high value and quality for building such models. By improving the models, the trade-off between the cost and benefit of digital agricultural systems can be improved, and incentives can be created to drive adoption.

\section{The role of data in digital agriculture}
\begin{figure}[ht]
	\centering
	\includegraphics[width=1\textwidth]{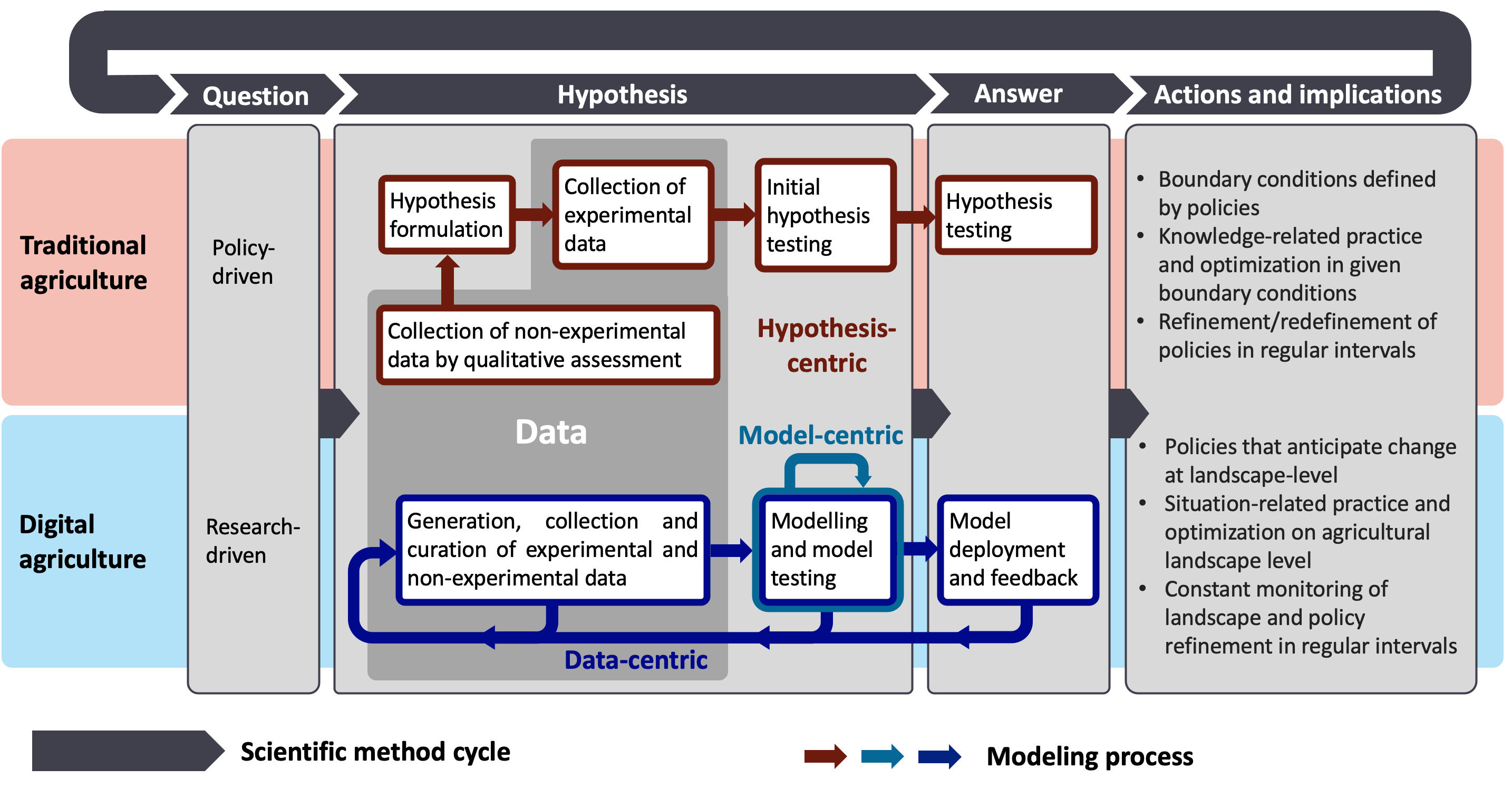}
	\caption{The scientific process in traditional and digital agriculture. The various types of modeling are highlighted in dark purple (data-centric), blue (model-centric), and red (hypothesis-centric). The scientific method cycle illustrates the feedback resulting from actions and implications to novel scientific questions.}
	\label{fig:data_driven_versus_hypothesis}
\end{figure} 

In digital agriculture, we observe that the significance of data quality is often underestimated, and the cascading effects of data quality throughout the entire digital agricultural ecosystem are frequently disregarded \cite{sambasivan2021everyone}. To define the criteria for high-quality data in this context, we elaborate on several key dimensions. However, it is important to note that these dimensions must still fully evolve and be proven in practice to have a clear understanding.
We characterize high-quality data in digital agriculture by diversity and completeness. It encompasses a wide range of samples and captures the nuances of different agricultural scenarios. This completeness ensures that all necessary information is available and strengthens the resilience of machine learning models by avoiding critical gaps.
Moreover, data accuracy is crucial as it measures how well the recorded values align with the true or expected values. Connected to this, we identify unbiasedness as necessary, which considers the presence or absence of systematic errors and consistent deviations. In the diverse landscape of agricultural data, for example, sampling bias happens when certain classes are overrepresented in the collection of data samples, such as data on diseases, which can result in distorted findings \cite{suresh2021framework}.
Further, we consider data consistency critical in digital agriculture to maintain the integrity of properties or attributes across different data sets. This is especially important when merging agricultural data sources to ensure seamless matching for different use cases.
Finally, we characterize high-quality data by its relevance to the intended agricultural task. It closely aligns with the requirements and context of the agricultural landscape, ensuring that models trained on such data are accurate and adaptable to different agricultural conditions.

In this section, we describe the use of data in traditional and digital agriculture and discuss three different paradigms in agriculture that differ in their modeling process and the integration and usage of data: hypothesis-centric traditional agriculture, model-centric digital agriculture, and data-centric digital agriculture (see Fig. \ref{fig:data_driven_versus_hypothesis}).

\textbf{Hypothesis-centric traditional agriculture.}
Traditional agriculture primarily adheres to a hypothesis-centric approach (red cycle in Fig. \ref{fig:data_driven_versus_hypothesis}), where farming practices are predominantly guided by qualitative assessments and observational data collected on-field. Farmers base their decision-making on long-established practices rooted in historical knowledge, non-experimental data, and qualitative assessments. This process is influenced by policies that are derived from experimental data. The data collected in experiments follow specific requirements for testing and evaluating a hypothesis. The objective in this context is often to test and validate existing agricultural hypotheses, seeking a deeper understanding of causal relationships within a particular environment.

\begin{figure}[ht]
	\centering
	\includegraphics[width=1\textwidth]{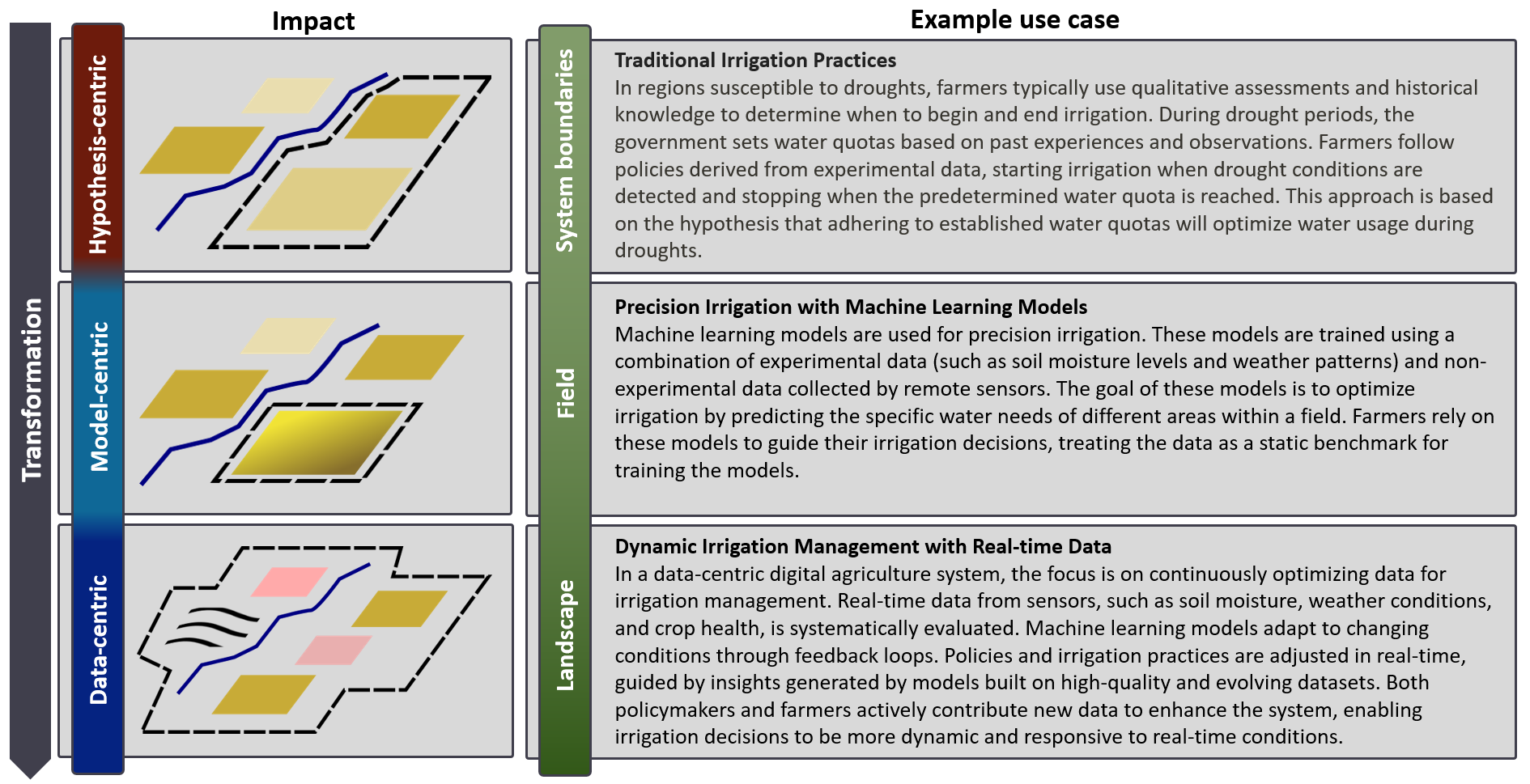}
	\caption{The transformative potential of data in agriculture. A shift from field and system boundaries to the landscape (green) can be achieved by moving from hypothesis-centric traditional agriculture to data-centric digital agriculture.}
	\label{fig:data_centric_agriculture}
\end{figure}

Actions in a policy-driven system can occur for two reasons. Firstly, practitioners may optimize their farming system within the existing system boundaries, such as reallocating limited resources. Secondly, a controlling instance may adjust ineffective policies and recommendations, redefining system boundaries. Consequently, implications in policy-driven systems are reactive, and proactive actions by farmers are not encouraged (Figure \ref{fig:data_centric_agriculture}, example 'System boundaries').

\textbf{Model-centric digital agriculture.} In digital agriculture, farmers' qualitative observations are complemented by non-experimental and experimental data collected by remote and proximal sensors. Depending on the application and the objectives, mechanistic, machine learning-based, or hybrid models are preferable \cite{baker2018mechanistic,ellis2020synergy,cohen2021dynamically}. Moreover, either non-experimental data only or both data types are used. So far, digital agriculture, particularly machine learning for agricultural applications, has predominantly been model-centric and focused on the design and evaluation of models \cite{sharma2020machine,shaikh2022towards} (blue cycle in Fig. \ref{fig:data_driven_versus_hypothesis}). These approaches primarily aim to enhance the accuracy and efficiency of the models themselves while treating the data as a static benchmark instead of a dynamic representation of the application. Despite the growing availability of agricultural data \cite{lu2020survey,gunder2022agricultural,kierdorf2023growliflower,wu2023extended} and advancements in methodologies, there is currently a point where many established machine learning methods achieve comparable levels of accuracy and face similar limitations in their applicability \cite{Maslej2023}.

While the common belief is that more data will lead to better models, building such models requires immense computational resources and leads to significant environmental impacts \cite{falk2023challenging}. However, limiting the size of datasets can lead to reduced generalization ability and incomplete information and insights. In the agricultural domain, plant development is influenced by complex interactions between genetic information and the environment. Given the vast number of potential environments and genotypes, and the lack of a well-established relationship between observable traits, genetic information, and the environment, it is crucial to have a comprehensive dataset that encompasses sufficient information. However, storing such a large amount of data is impractical due to the sheer volume of information generated in the digital agriculture domain. Storing every observed data point becomes increasingly unwieldy and may not align with the practical goals of data analysis. Essentially, trying to archive and analyze every piece of data will inevitably lead to an inefficient allocation of resources and computational efforts \cite{hadsell2020embracing}. 

A farmer's actions using model-centric approaches impact the fields to which they are applied. However, the models have no feedback loop regarding their impact at the field or farm levels. Consequently, the impacts in model-centric systems are confined to the field boundary, and any changes are expected to occur only within that boundary. Transformation at the landscape level is not promoted (Figure \ref{fig:data_centric_agriculture}, example 'Field').

\textbf{Data-centric digital agriculture.} Data-centric digital agriculture strongly emphasizes data, utilizing systematic and algorithmic approaches to identify and utilize rich and high-quality datasets for building models. This involves a comprehensive evaluation process to ensure that the model built on the dataset performs optimally for the intended task. 
Here, the data is optimized in a cycle, that means with several feedback loops and not treated as a static benchmark, but evolves jointly with the model development.
Data-centric approaches address the mentioned challenges of model-centric digital agriculture and identify valuable and high-quality data to build accurate, generalizable, and adaptable models \cite{sambasivan2021everyone,aroyo2022data}.  Therefore, data-centric digital agriculture is a logical next step to improve the widespread applicability and adoption of agricultural models.

Data-centric approaches have the potential to revolutionize the agricultural landscape by fostering the creation of accurate, generalizable, and adaptable models. By focusing on the data and its quality and recognizing and exploiting the dynamic nature of data in representing the agricultural environment, we can overcome the current limitations of machine learning models, which are restricted to small use cases and narrow application domains. Furthermore, the targeted use of high-quality data and data-centric approaches allows for a comprehensive analysis of the models and the detection of shortcomings, which can be corrected by constant feedback loops.
If policies permit, farmers can adjust their farming systems in real-time using predictions generated by a model built on high-quality data. Both policymakers and farmers can contribute newly collected data to the system, which enables insights to be gained across different systems and boundaries (see Figure \ref{fig:data_centric_agriculture}, example 'Landscape').

\section{Looking into the Future: A Data-centric Digital Agriculture} 

In this section, we explore the integration and use of data at every stage of the agricultural scientific process (see Fig. \ref{fig:data_driven_versus_hypothesis}) and highlight its central role in shaping agricultural research and practice. The section begins with data collection and curation, guided by the scientific question. It continues to model-centric agriculture, which includes important aspects such as designing models, optimizing parameters, and establishing evaluation metrics to assess model performance. Learning accurate models that generalize well and can adapt to new environments requires the process to be iterative, driven by insights and feedback gained from data and its interaction with the model. While these advancements can significantly improve the modeling process, they also present challenges, including the accessibility of data and sustainability aspects. Our elaborations will show the transformative potential of data-centric digital agriculture, providing insight into its crucial role in overcoming obstacles and shaping the future of agricultural science.

\subsection{Data collection and curation}
\subsubsection{Novel sensor and robotic technologies}
Sensing in agriculture often involves combining position sensors such as GNSS/GPS with other modalities, including remote sensing technologies like satellite or UAV imagery, environmental sensors that measure factors like temperature and humidity, and sophisticated instruments for assessing soil properties \cite{jin2021development,ahmadi2022towards}. The effectiveness of these modalities strongly depends on the specific quantity being estimated. While the shoot of a plant can be easily sensed using cameras, measuring solid properties, fluxes, genetic information, or root growth poses significant challenges. In many cases, only proxies can be measured, and the correlation to the desired quantity may be questioned. However, all of this information is relevant for describing the field's current state and making future predictions.

Especially in real field environments, current below-ground sensing systems have severe limitations. Often, it needs to be combined with domain knowledge of underlying processes and structures in the plant system, side-specific calibration, and complex soil models \cite{atkinson2019uncovering,yang2020crop}.
Similar limitations can be observed in some above-ground sensing techniques. Cameras with more spectral bands provide relevant information about plant traits, but hyperspectral sensing, particularly, still faces serious calibration issues when applied in the field under real-world conditions. Nevertheless, this technique is crucial for health monitoring, especially related to diseases \cite{wang2021review,saric2022applications}.

Integrating new, affordable, and powerful sensors, especially those with high spatial resolution and spectral capabilities, along with integrating different modalities, offers an opportunity for advanced applications in agriculture \cite{saiz-rubio2020,miranda2022controlled}. However, this potential still needs to be fully realized. One example is the combination of data from high-resolution above-ground sensors with low-resolution below-ground sensors. This combination allows for extracting more relevant traits, enabling more accurate modeling of the current situation. Additionally, this approach enables the semi-automatic generation of weakly labeled data across different modalities. One strategy to explore for this is transfer learning (see also \ref{sec:transfer}). Transfer learning involves leveraging knowledge gained from high-resolution sensors to enhance the performance of models using simpler, low-resolution sensors. For instance, high-resolution data can generate semi-automatic, weakly labeled data for low-resolution sensors. This concept is also connected to surrogate modeling, as the distilled knowledge from high-resolution sensors aids in training models for low-resolution counterparts.

Regarding robotics, several developments in agricultural operations will impact data acquisition and its utilization.
Firstly, robots for data acquisition are often aerial platforms that collect high-resolution data at frequent intervals \cite{finger2019,saiz-rubio2020,yang2020crop}. The collected phenotypical data provide valuable information for breeding applications and monitoring the current status quo \cite{yang2020crop,li2020review}.
Secondly, it is clear today that we will witness in the near future an increase in robots performing specific management tasks, such as weeding, often motivated by the desire to reduce agrochemical usage. By combining intervention actions like weeding with sensing and repeated operations, more detailed data can be obtained, allowing for the monitoring of treatment effectiveness on a larger scale \cite{busemeyer2013breedvision,schuster2023spatial,dewan2023development}.
Thirdly, robots have the potential to monitor and maintain field arrangements that are currently not considered at relevant scales, such as multi-cropping (patch-, strip-, inter-, pixel-cropping) \cite{donat2022,juventia2022spatio,hernandez2022model}. The usage of robots in these arrangements is expected to provide new data and insights.
Furthermore, the question of whether we will continue to see an increase in the size of agricultural machines or a shift towards fleets of smaller robots capable of performing multiple tasks is still under debate. However, using numerous small systems will impact data acquisition, as they will provide more distributed sensing \cite{shamshiri2018}.
Similar considerations apply to the development of robots that assist humans in specific operations, such as handling delicate fruits in orchards. In these cases, more detailed and process-related data can be captured.

\subsubsection{Active sensing}
Agricultural environments can be vast, cluttered, and filled with occlusions, making it difficult to sense relevant quantities of interest. Passive data recording has limitations as important information may not be directly observable or a large number of observations may be needed to cover fields or orchards densely. Therefore, robots need to change viewpoints, move to specific locations, or move obstacles to optimize data acquisition.
Active information gathering involves controlling a mobile system to obtain higher-quality observations that are more relevant to agricultural tasks or have higher accuracy. Existing approaches typically focus on optimizing sensor coverage or use sample-based viewpoint selection. Information-driven approaches are a fundamental solution but can be computationally challenging. For real-world applications, approximations are needed to make information-driven approaches applicable, e.g., for exploration \cite{stachniss2005rss}.

One promising approach to enhancing data acquisition in agriculture is through active sensing \cite{morrison2019multi,rehman2021viewpoint,stache2021adaptive,sun2023object}. Active sensing combines recording data with active learning and selective labeling paradigms \cite{foix2018task,wang2022unsupervised}. Selective labeling becomes particularly valuable in the dynamic and diverse agricultural landscape, where field and environmental conditions can vary significantly. It involves carefully choosing specific instances or features in the data for detailed annotation, which can significantly enhance the learning systems used for scene analysis in agriculture.
Further, the active sensing framework can employ techniques like uncertainty quantification methods \cite{Gawlikowski2023} and explainable machine learning \cite{roscher2020explainable}. These techniques provide valuable insights into the quality of the collected data and shed light on the decision-making processes of the models. By adopting approximative information-driven approaches that integrate data acquisition strategies and labeling needs, we can create models that are accurate and built on selectively collected, high-quality data. In the agricultural context, this means that the models can better adapt to the nuances of real-world conditions, leading to more effective and reliable outcomes in decision-making processes for farmers.

\subsubsection{Data generalization and specification}
\label{sec:gen}
In the field of data collection and curation for agriculture, the traditional 'one-size-fits-all' approach is already being replaced by a more nuanced strategy \cite{finger2019precision,shamshiri2018,lowenberg2015precision,saiz-rubio2020,carletto2021agricultural}. The traditional approach aimed to collect broad data that could be applied to various agricultural scenarios to achieve generality. However, this approach often lacked the specificity to address unique challenges in different regions or crops \cite{tseng2021timl,chen2022performance}. In contrast, the modern agricultural landscape requires more specific data, which is made possible by precision agriculture tools and sensors. Moreover, these technologies have become more accessible and affordable \cite{carletto2021agricultural}. Moving forward, it is necessary to balance the seemingly contrasting principles of generalization and specification.

To address this challenge, adaptable data strategies and, with them, adaptable models appear to be a promising approach. This transition requires a shift in how we view data - from a static, homogeneous resource to one that can be customized and refined to solve specific agricultural tasks. These strategies involve continuously gathering data and adjusting data collection methods and parameters to align with changing needs and tasks. They also rely on approaches that can utilize data and its quality information to adapt existing models for specific applications. However, several prerequisites must be met to effectively utilize adaptable data strategies.

Firstly, the data should be highly granular and have a clear structure that accurately reflects agricultural conditions. This allows for the extraction of detailed insights specific to different agricultural domains. Additionally, the data should be high-quality, meaning it should be complete, diverse, accurate, unbiased, consistent, and relevant, among other potential criteria \cite{sambasivan2021everyone,Whang2023}. Several methods show promise for creating (i.e., generating or collecting) and curating high-quality data. If a model is already available, it can be utilized to identify new data that improves and adapts the model (see Sec. \ref{sec:transfer}), as well as to detect less accurate or irrelevant data, thereby enhancing the quality of the dataset. If no model is available, incorporating data-only techniques such as data quality assessments \cite{carletto2021agricultural}, data sampling strategies \cite{kerry2010sampling,luo2019distributed}, data augmentation with real and synthetic data \cite{abbas2021tomato}, and data cleaning is crucial to ensure that the data remains a valuable resource for agricultural applications. The timeliness of the data and the ability to continuously update and refresh it are also essential requirements. Agricultural systems are dynamic, and the data should reflect these ongoing changes. Regular data updates keep models and decision support systems relevant and accurate. Here, edge computing, with its ability to process data near the source of data generation, can significantly reduce latency and enhance real-time decision support~\cite{zhang2020overview}.

Secondly, the data should be well-documented and tagged, providing metadata that describes its source, context, and quality \cite{koedel2022challenges}. Rich metadata is crucial for proper data curation and interpretation, as it helps ensuring data reliability and integrity. A significant development in this area is the emergence of foundation models, which are built on a vast amount of data from multiple domains and modalities, with the aim of adapting their knowledge to specific applications \cite{li2023foundation}. The transition towards adaptable data strategies, combined with foundation models, can potentially drive the next phase of digital agriculture, offering tailored solutions for various agricultural challenges. However, there are still open questions regarding their generalization capabilities, superiority over specialized models, ability to provide explanations and uncertainty measures, and inherent biases.

Furthermore, adaptable data strategies depend on open data standards and interoperability, which facilitate the integration of diverse datasets from various sources and formats \cite{koedel2022challenges}. This interoperability allows for the harmonious fusion and utilization of data from different regions, sensors, and crops to address unique challenges (see also Sec. \ref{sec:opendata}). 

\subsection{Modeling strategies and machine learning}

\subsubsection{Model transfer}
\label{sec:transfer}
In the dynamic agricultural environment, model transfer strategies provide exciting opportunities to foster adaptability \cite{autz2022pitfalls}. These techniques primarily focus on transferring knowledge and insights, often in the form of data, from one task or domain to another \cite{yang2020transfer}. By doing so, they maximize the use of existing models and minimize the need for extensive data collection in the target domain. 

Several approaches to transfer learning strategies can be employed in digital agriculture \cite{espejo2020towards,chen2022performance,al2022power,magistri2023one}. Instance-based transfer learning involves identifying relevant data samples from both domains that can be transferred to each other. This process allows models to adapt to and generalize better in new agricultural environments. Feature-based transfer learning focuses on migrating relevant features from one agricultural setting to another. For example, it can adapt features learned from one crop variety to enhance predictions for a different variety. Model-based transfer learning, often combined with a foundational model, involves using the knowledge of an existing model that is fine-tuned for a target application domain. Relation-based transfer learning can be used when shared relationships exist between agricultural systems. For instance, a model trained to optimize management strategies for one crop can leverage its understanding of environmental factors to benefit other crops, improving decision support and reducing the need for extensive, crop-specific data.

Acquiring high-quality, domain-specific data in the target task domain remains crucial for successful model transfer. In addition to collecting and curating new data, active learning and self-training strategies focus on a more efficient data acquisition \cite{doi:10.34133/icomputing.0058}. Active learning involves selecting the most informative data points, with human input for potential labeling. On the other hand, self-training allows the model to refine itself without human interaction, but it runs the risk of including inaccurate data. However, determining a suitable selection criterion for samples is still an ongoing research question \cite{tifrea2022uniform}.

To date, research in this field has mainly focused on deep machine learning, with a limited number of use cases. These investigations have primarily taken a model-centric approach, analyzing pre-trained model architectures originally designed for non-agricultural domains \cite{thenmozhi2019crop,espejo2020towards,chen2020using}. However, there are still important questions regarding the identification of valuable agricultural data for foundation models and data that can be effectively transferred. This challenge is amplified by the constraints of data collection in agricultural contexts, highlighting the need for precise and efficient data selection.

\subsubsection{Privacy-preserving models}
Integrating data-driven models, robotics, and emerging technologies such as edge computing in agriculture brings technological advancements and raises critical ethical considerations \cite{dara2022recommendations,Liang2022}. Privacy and fairness are among the primary ethical concerns that require careful attention during the development and deployment of these technologies.
Furthermore, it is necessary to ensure an unbiased data collection strategy and examine unbiased algorithms to ensure that they do not unintentionally disadvantage certain groups of farmers.

Privacy-preserving machine learning techniques have been developed to tackle these ethical concerns. These techniques protect sensitive data while enabling effective model training and inference. This is particularly important when working with agricultural data containing personal, financial, or proprietary information. Methods such as secure multi-party computation or federated learning can be employed to ensure data privacy in digital agriculture \cite{kumar2021sp2f,gardezi2023artificial,aggarwal2023federated}.

However, there is a trade-off between the benefits of open data and knowledge sharing and the need to protect sensitive agricultural data. The challenge is to find the proper balance or develop hybrid approaches that preserve privacy while advancing the field of digital agriculture. Striking this balance is a technical challenge and a moral and societal imperative as we navigate the ethical landscape of an increasingly digitized and data-driven agricultural sector. These considerations are particularly relevant in survey data collection, where maintaining the confidentiality and privacy of respondents is paramount.

\subsubsection{Sustainable machine learning}
Especially in the field of digital agriculture, sustainability is gaining recognition \cite{Walter2017,nations2020sustainable,khanna2022digital,basso2020digital}. 
As the global community becomes more aware of its critical importance, it is essential to consider not only how digital technologies can contribute to sustainable agriculture but also how these technologies can be designed sustainably \cite{van2021sustainable,falk2023challenging}.
In the computer vision and machine learning communities, promising directions are known as scalable machine learning, efficient machine learning, tiny machine learning, or frugal machine learning \cite{sanchez2020tinyml,al2015efficient}. These directions are of particular interest as they parallel the emergence of generally large foundation models.
In digital agriculture, sustainability has several facets and concerns the hardware (such as sensors, robots, low-power processors, and energy-efficient GPUs), the data, and the methods. Existing systems are mostly model-centric and focus on the energy efficiency and computational complexity of methods by designing them to be light-weight, transferable, scalable, and reusable \cite{mccool2017mixtures,li2019real,paudel2021machine}. Most works attempt to immediately handle the vast amount of data, treating it as static and disregarding the potential to guide model building through data collection and curation, including selection.
To design solutions that align with local agricultural and environmental contexts, interdisciplinary collaboration between agricultural experts, environmental scientists, data scientists, and machine learning engineers is necessary. By building on existing machine learning models, tailoring them effectively to the specific needs of a region or crop, and complementing the existing data with well-conceived data collection and curation, sustainability can be enhanced. This approach avoids the creation of complete datasets and the building of models from scratch. Generalization and specialization can be performed together, supporting sustainability (Sec. \ref{sec:gen}).
Furthermore, a core challenge is striking a balance between an open data culture and the responsible use of data. It is also crucial to ensure that the benefits of machine learning are accessible across different agricultural systems, from small-scale farms to large agribusinesses.

\subsection{Model evaluation and deployment}
\subsubsection{Evaluating the applicability of agricultural models}

Comprehensive evaluation strategies in agriculture are necessary to ensure the accuracy of models, but also generalizability and adaptability to assess the performance across various agricultural scenarios, encompassing different crops and environmental conditions~\cite{condran2022machine}. Previous research in this field has mainly focused on traditional machine learning evaluation metrics for accuracy, including overall accuracy, precision, recall, and F1 score. Additionally, metrics like mean absolute error and root mean square error have been used. 

So far, evaluating the generalization ability is mainly done by cross-validation with random partitions of a fixed data set into training and evaluation set \cite{a2019evaluation,du2022estimating}. The results on different evaluation sets are averaged. However, to understand a model's performance comprehensively, it is crucial to move beyond aggregate performance metrics and focus on fine-grained non-random subgroup analysis \cite{chung2019slice,koh2021wilds}. This involves evaluating the model's performance across attributes such as crop varieties, environmental factors, and geographical locations. Fine-grained subgroup evaluation provides deeper insights into the model's limitations and helps address potential biases or disparities in its performance.

In addition to subgroup analysis, evaluating the area of applicability of the model is becoming increasingly important \cite{meyer2021predicting}. Area of applicability refers to understanding where the model performs well and where its performance may decline. This includes testing the model on data points within and outside its training distribution. By systematically exploring the boundaries of the model's applicability with carefully designed evaluation sets, scenarios can be identified in which the model may exhibit uncertainty or lower performance. This approach is particularly valuable in digital agriculture, where diverse and evolving conditions necessitate a thorough understanding of a model's capabilities across various agricultural contexts.

Agricultural models, mainly those built on a limited number of labeled samples, face challenges associated with data shortcuts and spurious correlations that models may unintentionally rely on. Therefore, evaluation data should reflect diverse real-world settings to check whether the model's performance is not overly reliant on specific characteristics of the training data \cite{Liang2022}. If these characteristics are causally wrong, this is also known as the Clever Hans effect ~\cite{Schramowski2020a,kierdorf2023reliability}. Data ablation methods that systematically remove or alter portions of the input data to analyze the impact on a model's performance can be employed to identify potential model shortcuts and prevent the decision-making process from relying on such spurious correlations~\cite{kierdorf2023reliability}. 

In recent years, generative models have become powerful tools in digital agriculture. They can generate synthetic data, augment limited labeled datasets, and serve as data-driven growth models \cite{lu2022generative, drees2021temporal}. However, evaluating these models can be challenging due to the lack of reference data for direct comparison. Evaluation strategies used in the computer vision community may not be applicable in agriculture, as they have different objectives mostly related to perceptual quality. As a result, evaluation strategies for generative models often involve proxy tasks, such as plant size prediction or assessing the realism of generated images. With the increasing use of generative models in agriculture, it is crucial to develop robust evaluation methodologies that address the specific challenges of synthetic data generation. This will ensure the reliability and effectiveness of these models in diverse agricultural contexts.

\subsubsection{Uncertainty quantification and explainability machine learning}
In the dynamic landscape of digital agriculture, the reliability and trustworthiness of model results are of great importance for informed decision-support \cite{roscher2020explain,gardezi2023artificial}. Two critical elements in this field are explainability and uncertainty, which intersect and have significant implications for agricultural practices~\cite{loucks2005model}. Explainable machine learning combines domain knowledge with interpretation tools that present model components and the decision process of the model in a human-understandable way~\cite{roscher2020explain,kierdorf2023reliability}. At the same time, uncertainty encompasses potential variability in results due to uncertainties in agricultural data or knowledge gaps in the model~\cite{visser2021imprecision,Gawlikowski2023}.
Research in explainability and uncertainty offers transformative opportunities for robust model evaluation, a deeper understanding of the modeling processes, and actionable insights from agricultural data.
However, ensuring the reliability and calibration of uncertainty estimates in machine learning models remains a key challenge in data-centric digital agriculture~\cite{guo2017calibration, wilson2020bayesian}. 
Inaccurate uncertainty estimations make it challenging to identify out-of-domain scenarios, which means unknown environments~\cite{lee2018training} and lead to unexpected decisions~\cite{ovadia2019can, smith2018understanding}, creating a risk that is difficult to calculate.

While research has explored 'planning under uncertainty' and 'decision making under uncertainty', these approaches assume the existence of uncertainty estimates or estimate these uncertainties independently of explanations for the decisions and the origin of the uncertainties. Until now, explainable machine learning and uncertainty quantification have rarely been considered jointly. 
Moreover, to obtain reliable results for decision-support in digital agriculture, it is crucial to not only quantify uncertainty but also assess and calibrate the models~\cite{patel2021manifold, guo2017calibration}. Calibration of model outputs is also essential for meaningful interpretations and explanations that support agricultural interventions and decision support.
Efforts have been made to distinguish between model uncertainties and data uncertainties, which aids in identifying the sources of uncertainty~\cite{lakshminarayanan2017simple, kendall2017uncertainties}. However, combining explanations of specific decisions with information about the responsible components in the input and model can facilitate targeted model improvement~\cite{bau2020understanding}.
In digital agriculture, the iterative integration of uncertainty and explainability can guide the data collection process through active sensing and active labeling, enhancing the overall quality and usefulness of the collected data.

In this context, it is highly important that the expectations of the domain expert are clearly stated and the boundaries of possible explanations are well understood (see also confirmation bias,~\cite{adebayo2018sanity}). Furthermore, as already pointed out in \cite{roscher2020explainable}, it is mandatory for any application to have domain experts involved to arrive at domain-specific explanations and not just machine learning-derived interpretations.

\subsection{Fostering data accessibility and adoption}
\label{sec:opendata}
The FAIR principle is crucial in ensuring open access to high-quality research data. Data is considered FAIR when it is findable / discoverable, accessible, interoperable, and reusable. There is a growing trend among funding agencies, as demonstrated by the Bill \& Melinda Gates Foundation \cite{zotero-3810} and the Science for Humanity's Greatest Challenges (CGIAR, \cite{zotero-3813}), to require compliance with this principle.

However, the utility of a collected dataset or a trained model published in scientific papers might be limited to academic purposes and may not translate into practical applications. Bridging this gap demands additional efforts. A promising example can be drawn from the widespread availability of meteorological forecast data, such as that provided by the Deutscher Wetterdienst \cite{zotero-3814}, which can serve as a template for similar implementations in agricultural forecasting.

Beyond ensuring open access to data, the digitization of agricultural practices requires practitioners to be motivated and willing to adopt new technologies. This necessitates a concerted effort in policy development, farmer education \cite{ammann2022}, and showcasing the economic benefits of these technologies to the broader society \cite{ammann2022a}. Incorporating feedback loops into the agricultural landscape system, crucial for continuous model adaptation, relies on farmers sharing their data. Data sharing can either be facilitated by policies coupling this to subsidies or by farmers perceiving themselves as the primary beneficiaries of data sharing \cite{zhang2021}. Facilitation by policies requires creating a data exchange platform that works on the 'once-only' principle: The same data does not have to be entered manually multiple times. In many countries, farmers are required to enter their data into federal systems to receive subsidies essential for their operation. By creating an exchange platform that allows data exchange between several applications, digital possibilities can be used efficiently, and critical data can be requested from farmers without a high administrative burden. Beyond the implementation of such 'once-only' platforms, clearly defined data standards and measures to ensure high data quality need to be realized. Most importantly, clear rules are needed as to who has access to which data and what it may be used for. The development of these principles 
requires a great deal of knowledge on the side of policy-making federal or regional offices and cooperation with private industry to couple such solutions with standard farm-management systems. 

\section{Conclusion}
The future development and adoption of digital agriculture depend on several factors, with a key consideration being the balance between costs and benefits. To advance the field, a central goal is to develop agricultural models that are accurate, have a high generalization ability, and are adaptable to different agricultural systems and tasks. Achieving this ambitious goal requires advancements in data generation, collection, curation, and utilization through machine learning paradigms.

An often underestimated aspect in digital agriculture is the importance of data quality. High-quality data in this context is characterized by diversity, completeness, accuracy, unbiasedness, consistency, and relevance to the intended agricultural task. These dimensions highlight the need for a comprehensive understanding proven through practical experience. By prioritizing data quality, the digital agricultural ecosystem can strengthen machine learning models, preventing critical gaps and ensuring accurate alignment with true or expected values.

The paradigm of data-centric digital agriculture emerges as a promising next step, emphasizing the important role of data. Through systematic and algorithmic approaches, data-centric methodologies optimize datasets in a dynamic cycle, ensuring they evolve in parallel with model development. This differs from the current model-centric approach to digital agriculture, which treats data as a static benchmark and overlooks the dynamic nature of the agricultural environment. By addressing this challenge, data-centric strategies recognize and utilize valuable datasets, taking into account their dynamic nature. This fosters the development of accurate and transferable models that have a high generalization power.

Looking ahead, a data-centric approach has the potential to transform the agricultural landscape, fostering innovations in field arrangements and agricultural practices. With supportive policies, farmers can adapt their systems in real-time using predictions from models built on high-quality data. This reciprocal process, where both policy-makers and farmers contribute new data, facilitates insights across different systems and borders, demonstrating the potential of data-centric digital agriculture.

We argue that in the future of data-centric digital agriculture, key areas of focus include novel sensor and robotic technologies, active sensing strategies, data generalization and specification considerations, the use of model transfer strategies, privacy-preserving models, sustainable machine learning, a comprehensive evaluation of the applicability of agricultural models, and the use of uncertainty quantification and explainable machine learning. Another important aspect, although not discussed in this perspective, is the interdisciplinary collaboration between different communities and modelers. Therefore, data-centric digital agriculture does not imply exclusive reliance on data-driven models but instead emphasizes the need for action in this area.

\vspace{1em} \noindent\fbox{%
    \parbox{\textwidth}{%

\textbf{Box: The terminology of data}

Different types of data play a role in digital agriculture.

\emph{Raw observations} refer to the perception and recording of events, phenomena, or information with our senses and sensors. In agriculture, it is common to observe approximation or indirect indicators (proxies) rather than the actual phenomena of interest, such as biodiversity. This is due to practical limitations, feasibility constraints, and the complexity of directly measuring certain aspects in the field.\\

\emph{Data} in general refers to factual or numerical information that is processed for the purpose of analysis to produce meaningful insights or patterns to support decision-making. Data can be generated in many ways, such as by reproducing our knowledge, by using proximal and remote sensors, by computer simulations, or by instruments such as GPS devices. A data set can further contain labels or targets needed in the context of machine learning, in addition to the actual, processed raw observations.\\

In the area of agricultural data, there is a fundamental distinction between \emph{experimental data} and \emph{non-experimental data}, based on their origin~\cite{barnard1975}. Experimental data is systematically derived through controlled experiments designed to investigate cause-effect relationships while minimizing confounding factors~\cite{federer1955}. Examples include the thorough evaluation of new crop cultivars in variety testing before release~\cite{brown2020} or on-farm site-specific crop management experiments in precision agriculture~\cite{piepho2011}. These experiments follow standard design principles, including replications, randomization, and blocking.
On the other hand, non-experimental data stem from the natural and often uncontrolled settings of the agricultural landscape. This category encompasses a wide range, from coarse, aggregated data such as regional yield and quality values to high-resolution geospatial information collected by modern farming devices. Harvesters equipped with advanced sensors generate yield maps, while unmanned aerial vehicles (UAVs), satellites, and ground-based robots contribute geospatial information through remote and proximal sensing~\cite{kim2019unmanned,weiss2020remote,benami2021uniting}. Additionally, farmers' knowledge and survey data are considered non-experimental data.

Both experimental and non-experimental data play crucial roles in agricultural modeling, particularly in the training and testing of machine learning models, although with distinct objectives. Experimental data, due to their controlled nature, contribute to hypothesis validation and the understanding of causal relationships. The results enable targeted interventions and foster models that generalize well within specific parameters. Mechanistic models, pre-calibrated based on existing functional relationships between input and output data, are rooted in these controlled experiments.
On the other hand, non-experimental data provide a broader scope of observations, capturing the intricacies of real-world variability and complexity. These data, often collected due to the impracticality of setting up experiments in dynamic agricultural environments, serve as an invaluable resource for training machine learning models. These models excel at detecting patterns, correlations, and trends that naturally emerge. The synergy between experimental and non-experimental data is indispensable, offering comprehensive insights that bridge the gap between controlled experiments and the practical challenges of agriculture.\\

\emph{Data-driven methods} refer to approaches that use data and apply data analysis techniques, statistical modeling, and machine learning to extract patterns and insights for informed decision support.\\
    }
}

 \vspace{1em} 

\vspace{1em} \noindent\fbox{%
    \parbox{\textwidth}{%

\textbf{Box: The concurrent development of technologies shown by means of variable rate application of nitrogen}

In modern agriculture, N is applied in a site-specific manner with variable rate application (VRA) technology.
The spatial resolution of sensing techniques for VRA rapidly evolved over time, from manual soil sampling to automatic satellite-based sensing. 
Nowadays, the first generation of satellites provides open-access multi-spectral data in the tens of meter range (e.g., Landsat, \cite{wulder2022fifty}), and a new, second generation of satellites offers paywall-protected data in the meter to sub-meter range (e.g., Planet, SkySat).
Unmanned aerial vehicles (UAVs) offer another often localized, high-resolution, and high-quality data source for precision agriculture~\cite{argento2021}, provided that the basic concept of flight preparation and execution is taken into account~\cite{Roth2018a}.
While sensing technologies have reached sub-meter ranges, most modern farms still use fertilizer spreaders with working widths of 15 meters or more \cite{argento2021} to apply fertilizer uniformly across two or three management zones divided by farmers using farm-management information systems~\cite{Tummers2019}.

\vspace{2em}
\label{tab:history_precision_agriculture}
\footnotesize
\begin{tabular}{lllll}
	\toprule
	\multicolumn{2}{l}{Sensing technique} & Spatial resolution & Rate application level & Field of application \\
	\midrule
	\multicolumn{5}{l}{Traditional agriculture} \\
	\cmidrule{2-5}
	& Qualitative & Manual, field level  & Field level  & Generic \\
	\midrule
	\multicolumn{5}{l}{Modern agriculture} \\
	\cmidrule{2-5}
	 & Soil sampling & Manual, few sites & Site-specific ($\geq$ 50 m) & Very large fields \\
	& Remote sensing & Satellites ($\geq$ 30 m\textsuperscript{2}) & Tractor lane ($\geq$ 15 m) &  Very large fields \\
        & 1st generation \\
	& Remote sensing & Satellites (10 -- 0.7 m\textsuperscript{2}) & Tractor lane ($\geq$ 15 m) & Large fields \\
        & 2nd generation \\
	\midrule
	\multicolumn{5}{l}{Farm management systems and robotics} \\
	\cmidrule{2-5}
	& Proximal sensing & UAVs ($<$ 1 cm\textsuperscript{2}) & Tractor lane ($\geq$ 15 m) & Large fields\\
	 & Interactive sensing &  UAVs, robots ($<$ 1 cm\textsuperscript{2}) & Robot ($\leq$  1 cm) & Small fields (e.g., strips) \\
	\bottomrule
\end{tabular}

}}

 \vspace{1em}

\bmhead{Acknowledgments}

This work has partially been funded by the Deutsche Forschungsgemeinschaft (DFG, German Research Foundation) under Germany's Excellence Strategy, EXC-2070 - 390732324 - PhenoRob.
A.W. discloses support for the research of this work from Swiss National Science Foundation [grant number 200756]

\bibliography{references}


\begin{thebibliography}{133}
\ifx \bisbn   \undefined \def \bisbn  #1{ISBN #1}\fi
\ifx \binits  \undefined \def \binits#1{#1}\fi
\ifx \bauthor  \undefined \def \bauthor#1{#1}\fi
\ifx \batitle  \undefined \def \batitle#1{#1}\fi
\ifx \bjtitle  \undefined \def \bjtitle#1{#1}\fi
\ifx \bvolume  \undefined \def \bvolume#1{\textbf{#1}}\fi
\ifx \byear  \undefined \def \byear#1{#1}\fi
\ifx \bissue  \undefined \def \bissue#1{#1}\fi
\ifx \bfpage  \undefined \def \bfpage#1{#1}\fi
\ifx \blpage  \undefined \def \blpage #1{#1}\fi
\ifx \burl  \undefined \def \burl#1{\textsf{#1}}\fi
\ifx \doiurl  \undefined \def \doiurl#1{\url{https://doi.org/#1}}\fi
\ifx \betal  \undefined \def \betal{\textit{et al.}}\fi
\ifx \binstitute  \undefined \def \binstitute#1{#1}\fi
\ifx \binstitutionaled  \undefined \def \binstitutionaled#1{#1}\fi
\ifx \bctitle  \undefined \def \bctitle#1{#1}\fi
\ifx \beditor  \undefined \def \beditor#1{#1}\fi
\ifx \bpublisher  \undefined \def \bpublisher#1{#1}\fi
\ifx \bbtitle  \undefined \def \bbtitle#1{#1}\fi
\ifx \bedition  \undefined \def \bedition#1{#1}\fi
\ifx \bseriesno  \undefined \def \bseriesno#1{#1}\fi
\ifx \blocation  \undefined \def \blocation#1{#1}\fi
\ifx \bsertitle  \undefined \def \bsertitle#1{#1}\fi
\ifx \bsnm \undefined \def \bsnm#1{#1}\fi
\ifx \bsuffix \undefined \def \bsuffix#1{#1}\fi
\ifx \bparticle \undefined \def \bparticle#1{#1}\fi
\ifx \barticle \undefined \def \barticle#1{#1}\fi
\bibcommenthead
\ifx \bconfdate \undefined \def \bconfdate #1{#1}\fi
\ifx \botherref \undefined \def \botherref #1{#1}\fi
\ifx \url \undefined \def \url#1{\textsf{#1}}\fi
\ifx \bchapter \undefined \def \bchapter#1{#1}\fi
\ifx \bbook \undefined \def \bbook#1{#1}\fi
\ifx \bcomment \undefined \def \bcomment#1{#1}\fi
\ifx \oauthor \undefined \def \oauthor#1{#1}\fi
\ifx \citeauthoryear \undefined \def \citeauthoryear#1{#1}\fi
\ifx \endbibitem  \undefined \def \endbibitem {}\fi
\ifx \bconflocation  \undefined \def \bconflocation#1{#1}\fi
\ifx \arxivurl  \undefined \def \arxivurl#1{\textsf{#1}}\fi
\csname PreBibitemsHook\endcsname

\bibitem[\protect\citeauthoryear{Vermeulen et~al.}{2012}]{vermeulen2012climate}
\begin{barticle}
\bauthor{\bsnm{Vermeulen}, \binits{S.J.}},
\bauthor{\bsnm{Campbell}, \binits{B.M.}},
\bauthor{\bsnm{Ingram}, \binits{J.S.}}:
\batitle{Climate change and food systems}.
\bjtitle{Annual review of environment and resources}
\bvolume{37},
\bfpage{195}--\blpage{222}
(\byear{2012})
\end{barticle}
\endbibitem

\bibitem[\protect\citeauthoryear{Devot et~al.}{2023}]{Devot2023}
\begin{botherref}
\oauthor{\bsnm{Devot}, \binits{A.}},
\oauthor{\bsnm{Royer}, \binits{L.}},
\oauthor{\bsnm{{Caron Giauffret}}, \binits{E.}},
\oauthor{\bsnm{Ayral}, \binits{V.}}:
{The impact of extreme climate events on agricultural production in the EU}
(April)
(2023)
\end{botherref}
\endbibitem

\bibitem[\protect\citeauthoryear{Ortiz et~al.}{2021}]{ortiz2021review}
\begin{barticle}
\bauthor{\bsnm{Ortiz}, \binits{A.M.D.}},
\bauthor{\bsnm{Outhwaite}, \binits{C.L.}},
\bauthor{\bsnm{Dalin}, \binits{C.}},
\bauthor{\bsnm{Newbold}, \binits{T.}}:
\batitle{A review of the interactions between biodiversity, agriculture,
  climate change, and international trade: research and policy priorities}.
\bjtitle{One Earth}
\bvolume{4}(\bissue{1}),
\bfpage{88}--\blpage{101}
(\byear{2021})
\end{barticle}
\endbibitem

\bibitem[\protect\citeauthoryear{Walter et~al.}{2017}]{Walter2017}
\begin{barticle}
\bauthor{\bsnm{Walter}, \binits{A.}},
\bauthor{\bsnm{Finger}, \binits{R.}},
\bauthor{\bsnm{Huber}, \binits{R.}},
\bauthor{\bsnm{Buchmann}, \binits{N.}}:
\batitle{Opinion: {{Smart}} farming is key to developing sustainable
  agriculture}.
\bjtitle{Proceedings of the National Academy of Sciences}
\bvolume{114}(\bissue{24}),
\bfpage{6148}--\blpage{6150}
(\byear{2017})
\doiurl{10.1073/pnas.1707462114}
\end{barticle}
\endbibitem

\bibitem[\protect\citeauthoryear{Trendov et~al.}{2019}]{trendov2019digital}
\begin{botherref}
\oauthor{\bsnm{Trendov}, \binits{M.}},
\oauthor{\bsnm{Varas}, \binits{S.}},
\oauthor{\bsnm{Zeng}, \binits{M.}}, et al.:
Digital technologies in agriculture and rural areas: status report.
Digital technologies in agriculture and rural areas: status report.
(2019)
\end{botherref}
\endbibitem

\bibitem[\protect\citeauthoryear{Basso and Antle}{2020}]{basso2020digital}
\begin{barticle}
\bauthor{\bsnm{Basso}, \binits{B.}},
\bauthor{\bsnm{Antle}, \binits{J.}}:
\batitle{Digital agriculture to design sustainable agricultural systems}.
\bjtitle{Nature Sustainability}
\bvolume{3}(\bissue{4}),
\bfpage{254}--\blpage{256}
(\byear{2020})
\end{barticle}
\endbibitem

\bibitem[\protect\citeauthoryear{Kamilaris et~al.}{2017}]{kamilaris2017review}
\begin{barticle}
\bauthor{\bsnm{Kamilaris}, \binits{A.}},
\bauthor{\bsnm{Kartakoullis}, \binits{A.}},
\bauthor{\bsnm{Prenafeta-Bold{\'u}}, \binits{F.X.}}:
\batitle{A review on the practice of big data analysis in agriculture}.
\bjtitle{Computers and Electronics in Agriculture}
\bvolume{143},
\bfpage{23}--\blpage{37}
(\byear{2017})
\end{barticle}
\endbibitem

\bibitem[\protect\citeauthoryear{Coble et~al.}{2018}]{coble2018big}
\begin{barticle}
\bauthor{\bsnm{Coble}, \binits{K.H.}},
\bauthor{\bsnm{Mishra}, \binits{A.K.}},
\bauthor{\bsnm{Ferrell}, \binits{S.}},
\bauthor{\bsnm{Griffin}, \binits{T.}}:
\batitle{Big data in agriculture: A challenge for the future}.
\bjtitle{Applied Economic Perspectives and Policy}
\bvolume{40}(\bissue{1}),
\bfpage{79}--\blpage{96}
(\byear{2018})
\end{barticle}
\endbibitem

\bibitem[\protect\citeauthoryear{Shamshiri et~al.}{2018}]{shamshiri2018}
\begin{barticle}
\bauthor{\bsnm{Shamshiri}, \binits{R.R.}},
\bauthor{\bsnm{Weltzien}, \binits{C.}},
\bauthor{\bsnm{Hameed}, \binits{I.A.}},
\bauthor{\bsnm{Yule}, \binits{I.J.}},
\bauthor{\bsnm{Grift}, \binits{T.E.}},
\bauthor{\bsnm{Balasundram}, \binits{S.K.}},
\bauthor{\bsnm{Pitonakova}, \binits{L.}},
\bauthor{\bsnm{Ahmad}, \binits{D.}},
\bauthor{\bsnm{Chowdhary}, \binits{G.}}:
\batitle{Research and development in agricultural robotics: {{A}} perspective
  of digital farming}.
\bjtitle{International Journal of Agricultural and Biological Engineering}
\bvolume{11}(\bissue{4}),
\bfpage{1}--\blpage{14}
(\byear{2018})
\doiurl{10.25165/ijabe.v11i4.4278}
\end{barticle}
\endbibitem

\bibitem[\protect\citeauthoryear{Carletto
  et~al.}{2021}]{carletto2021agricultural}
\begin{bchapter}
\bauthor{\bsnm{Carletto}, \binits{C.}},
\bauthor{\bsnm{Dillon}, \binits{A.}},
\bauthor{\bsnm{Zezza}, \binits{A.}}:
\bctitle{Chapter 81 - agricultural data collection to minimize measurement
  error and maximize coverage}.
In: \beditor{\bsnm{Barrett}, \binits{C.B.}},
\beditor{\bsnm{Just}, \binits{D.R.}} (eds.)
\bbtitle{Handbook of Agricultural Economics}.
\bsertitle{Handbook of Agricultural Economics},
vol. \bseriesno{5},
pp. \bfpage{4407}--\blpage{4480}.
\bpublisher{Elsevier}, \blocation{???}
(\byear{2021}).
\doiurl{10.1016/bs.hesagr.2021.10.008} .
\burl{https://www.sciencedirect.com/science/article/pii/S1574007221000086}
\end{bchapter}
\endbibitem

\bibitem[\protect\citeauthoryear{Gebbers and Adamchuk}{2010}]{gebbers2010}
\begin{barticle}
\bauthor{\bsnm{Gebbers}, \binits{R.}},
\bauthor{\bsnm{Adamchuk}, \binits{V.I.}}:
\batitle{Precision {{Agriculture}} and {{Food Security}}}.
\bjtitle{Science}
\bvolume{327}(\bissue{5967}),
\bfpage{828}--\blpage{831}
(\byear{2010})
\doiurl{10.1126/science.1183899}
\end{barticle}
\endbibitem

\bibitem[\protect\citeauthoryear{Lowenberg-DeBoer}{2015}]{lowenberg2015precision}
\begin{barticle}
\bauthor{\bsnm{Lowenberg-DeBoer}, \binits{J.}}:
\batitle{The precision agriculture revolution}.
\bjtitle{Foreign Aff.}
\bvolume{94},
\bfpage{105}
(\byear{2015})
\end{barticle}
\endbibitem

\bibitem[\protect\citeauthoryear{Li et~al.}{2020}]{li2020review}
\begin{barticle}
\bauthor{\bsnm{Li}, \binits{Z.}},
\bauthor{\bsnm{Guo}, \binits{R.}},
\bauthor{\bsnm{Li}, \binits{M.}},
\bauthor{\bsnm{Chen}, \binits{Y.}},
\bauthor{\bsnm{Li}, \binits{G.}}:
\batitle{A review of computer vision technologies for plant phenotyping}.
\bjtitle{Computers and Electronics in Agriculture}
\bvolume{176},
\bfpage{105672}
(\byear{2020})
\end{barticle}
\endbibitem

\bibitem[\protect\citeauthoryear{Yang et~al.}{2020}]{yang2020crop}
\begin{barticle}
\bauthor{\bsnm{Yang}, \binits{W.}},
\bauthor{\bsnm{Feng}, \binits{H.}},
\bauthor{\bsnm{Zhang}, \binits{X.}},
\bauthor{\bsnm{Zhang}, \binits{J.}},
\bauthor{\bsnm{Doonan}, \binits{J.H.}},
\bauthor{\bsnm{Batchelor}, \binits{W.D.}},
\bauthor{\bsnm{Xiong}, \binits{L.}},
\bauthor{\bsnm{Yan}, \binits{J.}}:
\batitle{Crop phenomics and high-throughput phenotyping: past decades, current
  challenges, and future perspectives}.
\bjtitle{Molecular Plant}
\bvolume{13}(\bissue{2}),
\bfpage{187}--\blpage{214}
(\byear{2020})
\end{barticle}
\endbibitem

\bibitem[\protect\citeauthoryear{Sari{\'c}
  et~al.}{2022}]{saric2022applications}
\begin{botherref}
\oauthor{\bsnm{Sari{\'c}}, \binits{R.}},
\oauthor{\bsnm{Nguyen}, \binits{V.D.}},
\oauthor{\bsnm{Burge}, \binits{T.}},
\oauthor{\bsnm{Berkowitz}, \binits{O.}},
\oauthor{\bsnm{Trt{\'\i}lek}, \binits{M.}},
\oauthor{\bsnm{Whelan}, \binits{J.}},
\oauthor{\bsnm{Lewsey}, \binits{M.G.}},
\oauthor{\bsnm{{\v{C}}ustovi{\'c}}, \binits{E.}}:
Applications of hyperspectral imaging in plant phenotyping.
Trends in plant science
(2022)
\end{botherref}
\endbibitem

\bibitem[\protect\citeauthoryear{Hern{\'a}ndez-Ochoa
  et~al.}{2022}]{hernandez2022model}
\begin{barticle}
\bauthor{\bsnm{Hern{\'a}ndez-Ochoa}, \binits{I.M.}},
\bauthor{\bsnm{Gaiser}, \binits{T.}},
\bauthor{\bsnm{Kersebaum}, \binits{K.-C.}},
\bauthor{\bsnm{Webber}, \binits{H.}},
\bauthor{\bsnm{Seidel}, \binits{S.J.}},
\bauthor{\bsnm{Grahmann}, \binits{K.}},
\bauthor{\bsnm{Ewert}, \binits{F.}}:
\batitle{Model-based design of crop diversification through new field
  arrangements in spatially heterogeneous landscapes. a review}.
\bjtitle{Agronomy for Sustainable Development}
\bvolume{42}(\bissue{4}),
\bfpage{74}
(\byear{2022})
\end{barticle}
\endbibitem

\bibitem[\protect\citeauthoryear{Juventia et~al.}{2022}]{juventia2022spatio}
\begin{barticle}
\bauthor{\bsnm{Juventia}, \binits{S.D.}},
\bauthor{\bsnm{Nor{\'e}n}, \binits{I.L.S.}},
\bauthor{\bsnm{Van~Apeldoorn}, \binits{D.F.}},
\bauthor{\bsnm{Ditzler}, \binits{L.}},
\bauthor{\bsnm{Rossing}, \binits{W.A.}}:
\batitle{Spatio-temporal design of strip cropping systems}.
\bjtitle{Agricultural Systems}
\bvolume{201},
\bfpage{103455}
(\byear{2022})
\end{barticle}
\endbibitem

\bibitem[\protect\citeauthoryear{Tschumi et~al.}{2015}]{tschumi2015high}
\begin{bchapter}
\bauthor{\bsnm{Tschumi}, \binits{M.}},
\bauthor{\bsnm{Albrecht}, \binits{M.}},
\bauthor{\bsnm{Entling}, \binits{M.H.}},
\bauthor{\bsnm{Jacot}, \binits{K.}}:
\bctitle{{High effectiveness of tailored flower strips in reducing pests and
  crop plant damage}}.
In: \bbtitle{Proceedings of the Royal Society B: Biological Sciences},
vol. \bseriesno{282},
p. \bfpage{20151369}
(\byear{2015}).
\doiurl{10.1098/rspb.2015.1369} .
\burl{https://royalsocietypublishing.org/doi/10.1098/rspb.2015.1369}
\end{bchapter}
\endbibitem

\bibitem[\protect\citeauthoryear{Tscharntke
  et~al.}{2021}]{tscharntke2021beyond}
\begin{barticle}
\bauthor{\bsnm{Tscharntke}, \binits{T.}},
\bauthor{\bsnm{Grass}, \binits{I.}},
\bauthor{\bsnm{Wanger}, \binits{T.C.}},
\bauthor{\bsnm{Westphal}, \binits{C.}},
\bauthor{\bsnm{Bat{\'a}ry}, \binits{P.}}:
\batitle{Beyond organic farming--harnessing biodiversity-friendly landscapes}.
\bjtitle{Trends in ecology \& evolution}
\bvolume{36}(\bissue{10}),
\bfpage{919}--\blpage{930}
(\byear{2021})
\end{barticle}
\endbibitem

\bibitem[\protect\citeauthoryear{R{\"o}tter et~al.}{2015}]{rotter2015use}
\begin{barticle}
\bauthor{\bsnm{R{\"o}tter}, \binits{R.P.}},
\bauthor{\bsnm{Tao}, \binits{F.}},
\bauthor{\bsnm{H{\"o}hn}, \binits{J.G.}},
\bauthor{\bsnm{Palosuo}, \binits{T.}}:
\batitle{Use of crop simulation modelling to aid ideotype design of future
  cereal cultivars}.
\bjtitle{Journal of Experimental Botany}
\bvolume{66}(\bissue{12}),
\bfpage{3463}--\blpage{3476}
(\byear{2015})
\end{barticle}
\endbibitem

\bibitem[\protect\citeauthoryear{Muller and Martre}{2019}]{muller2019plant}
\begin{barticle}
\bauthor{\bsnm{Muller}, \binits{B.}},
\bauthor{\bsnm{Martre}, \binits{P.}}:
\batitle{Plant and crop simulation models: powerful tools to link physiology,
  genetics, and phenomics}.
\bjtitle{Journal of Experimental Botany}
\bvolume{70}(\bissue{9}),
\bfpage{2339}--\blpage{2344}
(\byear{2019})
\end{barticle}
\endbibitem

\bibitem[\protect\citeauthoryear{Drees et~al.}{2021}]{drees2021temporal}
\begin{barticle}
\bauthor{\bsnm{Drees}, \binits{L.}},
\bauthor{\bsnm{Junker-Frohn}, \binits{L.V.}},
\bauthor{\bsnm{Kierdorf}, \binits{J.}},
\bauthor{\bsnm{Roscher}, \binits{R.}}:
\batitle{Temporal prediction and evaluation of brassica growth in the field
  using conditional generative adversarial networks}.
\bjtitle{Computers and Electronics in Agriculture}
\bvolume{190},
\bfpage{106415}
(\byear{2021})
\end{barticle}
\endbibitem

\bibitem[\protect\citeauthoryear{Miranda et~al.}{2022}]{miranda2022controlled}
\begin{bchapter}
\bauthor{\bsnm{Miranda}, \binits{M.}},
\bauthor{\bsnm{Drees}, \binits{L.}},
\bauthor{\bsnm{Roscher}, \binits{R.}}:
\bctitle{Controlled multi-modal image generation for plant growth modeling}.
In: \bbtitle{Proc. of the International Conference on Pattern Recognition
  (ICPR)},
pp. \bfpage{5118}--\blpage{5124}
(\byear{2022}).
\bcomment{IEEE}
\end{bchapter}
\endbibitem

\bibitem[\protect\citeauthoryear{Donat et~al.}{2022}]{donat2022}
\begin{barticle}
\bauthor{\bsnm{Donat}, \binits{M.}},
\bauthor{\bsnm{Geistert}, \binits{J.}},
\bauthor{\bsnm{Grahmann}, \binits{K.}},
\bauthor{\bsnm{Bloch}, \binits{R.}},
\bauthor{\bsnm{{Bellingrath-Kimura}}, \binits{S.D.}}:
\batitle{Patch cropping- a new methodological approach to determine new field
  arrangements that increase the multifunctionality of agricultural
  landscapes}.
\bjtitle{Computers and Electronics in Agriculture}
\bvolume{197},
\bfpage{106894}
(\byear{2022})
\doiurl{10.1016/j.compag.2022.106894}
\end{barticle}
\endbibitem

\bibitem[\protect\citeauthoryear{Perich et~al.}{2023}]{perich2023pixel}
\begin{barticle}
\bauthor{\bsnm{Perich}, \binits{G.}},
\bauthor{\bsnm{Turkoglu}, \binits{M.O.}},
\bauthor{\bsnm{Graf}, \binits{L.V.}},
\bauthor{\bsnm{Wegner}, \binits{J.D.}},
\bauthor{\bsnm{Aasen}, \binits{H.}},
\bauthor{\bsnm{Walter}, \binits{A.}},
\bauthor{\bsnm{Liebisch}, \binits{F.}}:
\batitle{Pixel-based yield mapping and prediction from sentinel-2 using
  spectral indices and neural networks}.
\bjtitle{Field Crops Research}
\bvolume{292},
\bfpage{108824}
(\byear{2023})
\end{barticle}
\endbibitem

\bibitem[\protect\citeauthoryear{Finger et~al.}{2019}]{finger2019precision}
\begin{barticle}
\bauthor{\bsnm{Finger}, \binits{R.}},
\bauthor{\bsnm{Swinton}, \binits{S.M.}},
\bauthor{\bsnm{El~Benni}, \binits{N.}},
\bauthor{\bsnm{Walter}, \binits{A.}}:
\batitle{Precision farming at the nexus of agricultural production and the
  environment}.
\bjtitle{Annual Review of Resource Economics}
\bvolume{11},
\bfpage{313}--\blpage{335}
(\byear{2019})
\end{barticle}
\endbibitem

\bibitem[\protect\citeauthoryear{Weyler et~al.}{2021}]{weyler2021joint}
\begin{barticle}
\bauthor{\bsnm{Weyler}, \binits{J.}},
\bauthor{\bsnm{Milioto}, \binits{A.}},
\bauthor{\bsnm{Falck}, \binits{T.}},
\bauthor{\bsnm{Behley}, \binits{J.}},
\bauthor{\bsnm{Stachniss}, \binits{C.}}:
\batitle{Joint plant instance detection and leaf count estimation for in-field
  plant phenotyping}.
\bjtitle{IEEE Robotics and Automation Letters}
\bvolume{6}(\bissue{2}),
\bfpage{3599}--\blpage{3606}
(\byear{2021})
\end{barticle}
\endbibitem

\bibitem[\protect\citeauthoryear{Marks et~al.}{2022}]{marks2022precise}
\begin{bchapter}
\bauthor{\bsnm{Marks}, \binits{E.}},
\bauthor{\bsnm{Magistri}, \binits{F.}},
\bauthor{\bsnm{Stachniss}, \binits{C.}}:
\bctitle{Precise 3d reconstruction of plants from uav imagery combining bundle
  adjustment and template matching}.
In: \bbtitle{Proc. of the International Conference on Robotics and Automation
  (ICRA)},
pp. \bfpage{2259}--\blpage{2265}
(\byear{2022})
\end{bchapter}
\endbibitem

\bibitem[\protect\citeauthoryear{Shaikh et~al.}{2022}]{shaikh2022towards}
\begin{barticle}
\bauthor{\bsnm{Shaikh}, \binits{T.A.}},
\bauthor{\bsnm{Rasool}, \binits{T.}},
\bauthor{\bsnm{Lone}, \binits{F.R.}}:
\batitle{Towards leveraging the role of machine learning and artificial
  intelligence in precision agriculture and smart farming}.
\bjtitle{Computers and Electronics in Agriculture}
\bvolume{198},
\bfpage{107119}
(\byear{2022})
\end{barticle}
\endbibitem

\bibitem[\protect\citeauthoryear{Lottes et~al.}{2018}]{lottes2018}
\begin{barticle}
\bauthor{\bsnm{Lottes}, \binits{P.}},
\bauthor{\bsnm{Behley}, \binits{J.}},
\bauthor{\bsnm{Milioto}, \binits{A.}},
\bauthor{\bsnm{Stachniss}, \binits{C.}}:
\batitle{Fully convolutional networks with sequential information for robust
  crop and weed detection in precision farming}.
\bjtitle{IEEE Robotics and Automation Letters}
\bvolume{3}(\bissue{4}),
\bfpage{2870}--\blpage{2877}
(\byear{2018})
\doiurl{10.1109/LRA.2018.2846289}
\end{barticle}
\endbibitem

\bibitem[\protect\citeauthoryear{Pretto et~al.}{2021}]{pretto2021}
\begin{barticle}
\bauthor{\bsnm{Pretto}, \binits{A.}},
\bauthor{\bsnm{Aravecchia}, \binits{S.}},
\bauthor{\bsnm{Burgard}, \binits{W.}},
\bauthor{\bsnm{Chebrolu}, \binits{N.}},
\bauthor{\bsnm{Dornhege}, \binits{C.}},
\bauthor{\bsnm{Falck}, \binits{T.}},
\bauthor{\bsnm{Fleckenstein}, \binits{F.}},
\bauthor{\bsnm{Fontenla}, \binits{A.}},
\bauthor{\bsnm{Imperoli}, \binits{M.}},
\bauthor{\bsnm{Khanna}, \binits{R.}},
\bauthor{\bsnm{Liebisch}, \binits{F.}},
\bauthor{\bsnm{Lottes}, \binits{P.}},
\bauthor{\bsnm{Milioto}, \binits{A.}},
\bauthor{\bsnm{Nardi}, \binits{D.}},
\bauthor{\bsnm{Nardi}, \binits{S.}},
\bauthor{\bsnm{Pfeifer}, \binits{J.}},
\bauthor{\bsnm{Popovi{\'c}}, \binits{M.}},
\bauthor{\bsnm{Potena}, \binits{C.}},
\bauthor{\bsnm{Pradalier}, \binits{C.}},
\bauthor{\bsnm{{Rothacker-Feder}}, \binits{E.}},
\bauthor{\bsnm{Sa}, \binits{I.}},
\bauthor{\bsnm{Schaefer}, \binits{A.}},
\bauthor{\bsnm{Siegwart}, \binits{R.}},
\bauthor{\bsnm{Stachniss}, \binits{C.}},
\bauthor{\bsnm{Walter}, \binits{A.}},
\bauthor{\bsnm{Winterhalter}, \binits{W.}},
\bauthor{\bsnm{Wu}, \binits{X.}},
\bauthor{\bsnm{Nieto}, \binits{J.}}:
\batitle{Building an {{Aerial}}\textendash{{Ground Robotics System}} for
  {{Precision Farming}}: {{An Adaptable Solution}}}.
\bjtitle{IEEE Robotics \& Automation Magazine}
\bvolume{28}(\bissue{3}),
\bfpage{29}--\blpage{49}
(\byear{2021})
\doiurl{10.1109/MRA.2020.3012492}
\end{barticle}
\endbibitem

\bibitem[\protect\citeauthoryear{Sambasivan
  et~al.}{2021}]{sambasivan2021everyone}
\begin{bchapter}
\bauthor{\bsnm{Sambasivan}, \binits{N.}},
\bauthor{\bsnm{Kapania}, \binits{S.}},
\bauthor{\bsnm{Highfill}, \binits{H.}},
\bauthor{\bsnm{Akrong}, \binits{D.}},
\bauthor{\bsnm{Paritosh}, \binits{P.}},
\bauthor{\bsnm{Aroyo}, \binits{L.M.}}:
\bctitle{“everyone wants to do the model work, not the data work”: Data
  cascades in high-stakes ai}.
In: \bbtitle{Proc. of the CHI Conference on Human Factors in Computing
  Systems},
pp. \bfpage{1}--\blpage{15}
(\byear{2021})
\end{bchapter}
\endbibitem

\bibitem[\protect\citeauthoryear{Suresh and Guttag}{2021}]{suresh2021framework}
\begin{bchapter}
\bauthor{\bsnm{Suresh}, \binits{H.}},
\bauthor{\bsnm{Guttag}, \binits{J.}}:
\bctitle{A framework for understanding sources of harm throughout the machine
  learning life cycle}.
In: \bbtitle{Proceedings of the 1st ACM Conference on Equity and Access in
  Algorithms, Mechanisms, and Optimization}.
\bsertitle{EAAMO '21}.
\bpublisher{Association for Computing Machinery},
\blocation{New York, NY, USA}
(\byear{2021}).
\doiurl{10.1145/3465416.3483305} .
\burl{https://doi.org/10.1145/3465416.3483305}
\end{bchapter}
\endbibitem

\bibitem[\protect\citeauthoryear{Baker et~al.}{2018}]{baker2018mechanistic}
\begin{barticle}
\bauthor{\bsnm{Baker}, \binits{R.E.}},
\bauthor{\bsnm{Pena}, \binits{J.-M.}},
\bauthor{\bsnm{Jayamohan}, \binits{J.}},
\bauthor{\bsnm{J{\'e}rusalem}, \binits{A.}}:
\batitle{Mechanistic models versus machine learning, a fight worth fighting for
  the biological community?}
\bjtitle{Biology letters}
\bvolume{14}(\bissue{5}),
\bfpage{20170660}
(\byear{2018})
\end{barticle}
\endbibitem

\bibitem[\protect\citeauthoryear{Ellis et~al.}{2020}]{ellis2020synergy}
\begin{barticle}
\bauthor{\bsnm{Ellis}, \binits{J.}},
\bauthor{\bsnm{Jacobs}, \binits{M.}},
\bauthor{\bsnm{Dijkstra}, \binits{J.}},
\bauthor{\bsnm{Laar}, \binits{H.}},
\bauthor{\bsnm{Cant}, \binits{J.}},
\bauthor{\bsnm{Tulpan}, \binits{D.}},
\bauthor{\bsnm{Ferguson}, \binits{N.}}:
\batitle{Synergy between mechanistic modelling and data-driven models for
  modern animal production systems in the era of big data}.
\bjtitle{Animal}
\bvolume{14}(\bissue{S2}),
\bfpage{223}--\blpage{237}
(\byear{2020})
\end{barticle}
\endbibitem

\bibitem[\protect\citeauthoryear{Cohen et~al.}{2021}]{cohen2021dynamically}
\begin{barticle}
\bauthor{\bsnm{Cohen}, \binits{A.R.}},
\bauthor{\bsnm{Chen}, \binits{G.}},
\bauthor{\bsnm{Berger}, \binits{E.M.}},
\bauthor{\bsnm{Warrier}, \binits{S.}},
\bauthor{\bsnm{Lan}, \binits{G.}},
\bauthor{\bsnm{Grubert}, \binits{E.}},
\bauthor{\bsnm{Dellaert}, \binits{F.}},
\bauthor{\bsnm{Chen}, \binits{Y.}}:
\batitle{Dynamically controlled environment agriculture: Integrating machine
  learning and mechanistic and physiological models for sustainable food
  cultivation}.
\bjtitle{ACS ES\&T Engineering}
\bvolume{2}(\bissue{1}),
\bfpage{3}--\blpage{19}
(\byear{2021})
\end{barticle}
\endbibitem

\bibitem[\protect\citeauthoryear{Sharma et~al.}{2020}]{sharma2020machine}
\begin{barticle}
\bauthor{\bsnm{Sharma}, \binits{A.}},
\bauthor{\bsnm{Jain}, \binits{A.}},
\bauthor{\bsnm{Gupta}, \binits{P.}},
\bauthor{\bsnm{Chowdary}, \binits{V.}}:
\batitle{Machine learning applications for precision agriculture: A
  comprehensive review}.
\bjtitle{IEEE Access}
\bvolume{9},
\bfpage{4843}--\blpage{4873}
(\byear{2020})
\end{barticle}
\endbibitem

\bibitem[\protect\citeauthoryear{Lu and Young}{2020}]{lu2020survey}
\begin{barticle}
\bauthor{\bsnm{Lu}, \binits{Y.}},
\bauthor{\bsnm{Young}, \binits{S.}}:
\batitle{A survey of public datasets for computer vision tasks in precision
  agriculture}.
\bjtitle{Computers and Electronics in Agriculture}
\bvolume{178},
\bfpage{105760}
(\byear{2020})
\end{barticle}
\endbibitem

\bibitem[\protect\citeauthoryear{G{\"u}nder
  et~al.}{2022}]{gunder2022agricultural}
\begin{barticle}
\bauthor{\bsnm{G{\"u}nder}, \binits{M.}},
\bauthor{\bsnm{Ispizua~Yamati}, \binits{F.R.}},
\bauthor{\bsnm{Kierdorf}, \binits{J.}},
\bauthor{\bsnm{Roscher}, \binits{R.}},
\bauthor{\bsnm{Mahlein}, \binits{A.-K.}},
\bauthor{\bsnm{Bauckhage}, \binits{C.}}:
\batitle{Agricultural plant cataloging and establishment of a data framework
  from uav-based crop images by computer vision}.
\bjtitle{GigaScience}
\bvolume{11},
\bfpage{054}
(\byear{2022})
\end{barticle}
\endbibitem

\bibitem[\protect\citeauthoryear{Kierdorf
  et~al.}{2023}]{kierdorf2023growliflower}
\begin{barticle}
\bauthor{\bsnm{Kierdorf}, \binits{J.}},
\bauthor{\bsnm{Junker-Frohn}, \binits{L.V.}},
\bauthor{\bsnm{Delaney}, \binits{M.}},
\bauthor{\bsnm{Olave}, \binits{M.D.}},
\bauthor{\bsnm{Burkart}, \binits{A.}},
\bauthor{\bsnm{Jaenicke}, \binits{H.}},
\bauthor{\bsnm{Muller}, \binits{O.}},
\bauthor{\bsnm{Rascher}, \binits{U.}},
\bauthor{\bsnm{Roscher}, \binits{R.}}:
\batitle{Growliflower: An image time-series dataset for growth analysis of
  cauliflower}.
\bjtitle{Journal of Field Robotics}
\bvolume{40}(\bissue{2}),
\bfpage{173}--\blpage{192}
(\byear{2023})
\end{barticle}
\endbibitem

\bibitem[\protect\citeauthoryear{Wu et~al.}{2023}]{wu2023extended}
\begin{botherref}
\oauthor{\bsnm{Wu}, \binits{J.}},
\oauthor{\bsnm{Pichler}, \binits{D.}},
\oauthor{\bsnm{Marley}, \binits{D.}},
\oauthor{\bsnm{Hovakimyan}, \binits{N.}},
\oauthor{\bsnm{Wilson}, \binits{D.A.}},
\oauthor{\bsnm{Hobbs}, \binits{J.}}:
Extended agriculture-vision: An extension of a large aerial image dataset for
  agricultural pattern analysis.
Transactions on Machine Learning Research
(2023)
\end{botherref}
\endbibitem

\bibitem[\protect\citeauthoryear{Maslej et~al.}{2023}]{Maslej2023}
\begin{botherref}
\oauthor{\bsnm{Maslej}, \binits{N.}},
\oauthor{\bsnm{Fattorini}, \binits{L.}},
\oauthor{\bsnm{Brynjolfsson}, \binits{E.}},
\oauthor{\bsnm{Etchemendy}, \binits{J.}},
\oauthor{\bsnm{Ligett}, \binits{K.}},
\oauthor{\bsnm{Lyons}, \binits{T.}},
\oauthor{\bsnm{Manyika}, \binits{J.}},
\oauthor{\bsnm{Ngo}, \binits{H.}},
\oauthor{\bsnm{Niebles}, \binits{J.C.}},
\oauthor{\bsnm{Parli}, \binits{V.}},
\oauthor{\bsnm{Shoham}, \binits{Y.}},
\oauthor{\bsnm{Wald}, \binits{R.}},
\oauthor{\bsnm{Clark}, \binits{J.}},
\oauthor{\bsnm{Perrault}, \binits{R.}}:
{The AI Index 2023 Annual Report}
(2023)
\end{botherref}
\endbibitem

\bibitem[\protect\citeauthoryear{Falk and van
  Wynsberghe}{2023}]{falk2023challenging}
\begin{botherref}
\oauthor{\bsnm{Falk}, \binits{S.}},
\oauthor{\bsnm{Wynsberghe}, \binits{A.}}:
Challenging ai for sustainability: what ought it mean?
AI and Ethics,
1--11
(2023)
\end{botherref}
\endbibitem

\bibitem[\protect\citeauthoryear{Hadsell et~al.}{2020}]{hadsell2020embracing}
\begin{barticle}
\bauthor{\bsnm{Hadsell}, \binits{R.}},
\bauthor{\bsnm{Rao}, \binits{D.}},
\bauthor{\bsnm{Rusu}, \binits{A.A.}},
\bauthor{\bsnm{Pascanu}, \binits{R.}}:
\batitle{Embracing change: Continual learning in deep neural networks}.
\bjtitle{Trends in cognitive sciences}
\bvolume{24}(\bissue{12}),
\bfpage{1028}--\blpage{1040}
(\byear{2020})
\end{barticle}
\endbibitem

\bibitem[\protect\citeauthoryear{Aroyo et~al.}{2022}]{aroyo2022data}
\begin{barticle}
\bauthor{\bsnm{Aroyo}, \binits{L.}},
\bauthor{\bsnm{Lease}, \binits{M.}},
\bauthor{\bsnm{Paritosh}, \binits{P.}},
\bauthor{\bsnm{Schaekermann}, \binits{M.}}:
\batitle{Data excellence for ai: why should you care?}
\bjtitle{Interactions}
\bvolume{29}(\bissue{2}),
\bfpage{66}--\blpage{69}
(\byear{2022})
\end{barticle}
\endbibitem

\bibitem[\protect\citeauthoryear{Jin et~al.}{2021}]{jin2021development}
\begin{barticle}
\bauthor{\bsnm{Jin}, \binits{Y.}},
\bauthor{\bsnm{Liu}, \binits{J.}},
\bauthor{\bsnm{Xu}, \binits{Z.}},
\bauthor{\bsnm{Yuan}, \binits{S.}},
\bauthor{\bsnm{Li}, \binits{P.}},
\bauthor{\bsnm{Wang}, \binits{J.}}:
\batitle{Development status and trend of agricultural robot technology}.
\bjtitle{International Journal of Agricultural and Biological Engineering}
\bvolume{14}(\bissue{4}),
\bfpage{1}--\blpage{19}
(\byear{2021})
\end{barticle}
\endbibitem

\bibitem[\protect\citeauthoryear{Ahmadi et~al.}{2022}]{ahmadi2022towards}
\begin{bchapter}
\bauthor{\bsnm{Ahmadi}, \binits{A.}},
\bauthor{\bsnm{Halstead}, \binits{M.}},
\bauthor{\bsnm{McCool}, \binits{C.}}:
\bctitle{Towards autonomous visual navigation in arable fields}.
In: \bbtitle{Proc. of the IEEE/RSJ International Conference on Intelligent
  Robots and Systems (IROS)},
pp. \bfpage{6585}--\blpage{6592}
(\byear{2022})
\end{bchapter}
\endbibitem

\bibitem[\protect\citeauthoryear{Atkinson
  et~al.}{2019}]{atkinson2019uncovering}
\begin{barticle}
\bauthor{\bsnm{Atkinson}, \binits{J.A.}},
\bauthor{\bsnm{Pound}, \binits{M.P.}},
\bauthor{\bsnm{Bennett}, \binits{M.J.}},
\bauthor{\bsnm{Wells}, \binits{D.M.}}:
\batitle{Uncovering the hidden half of plants using new advances in root
  phenotyping}.
\bjtitle{Current Opinion in Biotechnology}
\bvolume{55},
\bfpage{1}--\blpage{8}
(\byear{2019})
\end{barticle}
\endbibitem

\bibitem[\protect\citeauthoryear{Wang et~al.}{2021}]{wang2021review}
\begin{barticle}
\bauthor{\bsnm{Wang}, \binits{C.}},
\bauthor{\bsnm{Liu}, \binits{B.}},
\bauthor{\bsnm{Liu}, \binits{L.}},
\bauthor{\bsnm{Zhu}, \binits{Y.}},
\bauthor{\bsnm{Hou}, \binits{J.}},
\bauthor{\bsnm{Liu}, \binits{P.}},
\bauthor{\bsnm{Li}, \binits{X.}}:
\batitle{A review of deep learning used in the hyperspectral image analysis for
  agriculture}.
\bjtitle{Artificial Intelligence Review}
\bvolume{54}(\bissue{7}),
\bfpage{5205}--\blpage{5253}
(\byear{2021})
\end{barticle}
\endbibitem

\bibitem[\protect\citeauthoryear{{Saiz-Rubio} and
  {Rovira-M{\'a}s}}{2020}]{saiz-rubio2020}
\begin{barticle}
\bauthor{\bsnm{{Saiz-Rubio}}, \binits{V.}},
\bauthor{\bsnm{{Rovira-M{\'a}s}}, \binits{F.}}:
\batitle{From {{Smart Farming}} towards {{Agriculture}} 5.0: {{A Review}} on
  {{Crop Data Management}}}.
\bjtitle{Agronomy}
\bvolume{10}(\bissue{2}),
\bfpage{207}
(\byear{2020})
\doiurl{10.3390/agronomy10020207}
\end{barticle}
\endbibitem

\bibitem[\protect\citeauthoryear{Finger et~al.}{2019}]{finger2019}
\begin{barticle}
\bauthor{\bsnm{Finger}, \binits{R.}},
\bauthor{\bsnm{Swinton}, \binits{S.M.}},
\bauthor{\bsnm{El~Benni}, \binits{N.}},
\bauthor{\bsnm{Walter}, \binits{A.}}:
\batitle{Precision {{Farming}} at the {{Nexus}} of {{Agricultural Production}}
  and the {{Environment}}}.
\bjtitle{Annual Review of Resource Economics}
\bvolume{11}(\bissue{1}),
\bfpage{313}--\blpage{335}
(\byear{2019})
\doiurl{10.1146/annurev-resource-100518-093929}
\end{barticle}
\endbibitem

\bibitem[\protect\citeauthoryear{Busemeyer
  et~al.}{2013}]{busemeyer2013breedvision}
\begin{barticle}
\bauthor{\bsnm{Busemeyer}, \binits{L.}},
\bauthor{\bsnm{Mentrup}, \binits{D.}},
\bauthor{\bsnm{M{\"o}ller}, \binits{K.}},
\bauthor{\bsnm{Wunder}, \binits{E.}},
\bauthor{\bsnm{Alheit}, \binits{K.}},
\bauthor{\bsnm{Hahn}, \binits{V.}},
\bauthor{\bsnm{Maurer}, \binits{H.P.}},
\bauthor{\bsnm{Reif}, \binits{J.C.}},
\bauthor{\bsnm{W{\"u}rschum}, \binits{T.}},
\bauthor{\bsnm{M{\"u}ller}, \binits{J.}}, \betal:
\batitle{Breedvision—a multi-sensor platform for non-destructive field-based
  phenotyping in plant breeding}.
\bjtitle{Sensors}
\bvolume{13}(\bissue{3}),
\bfpage{2830}--\blpage{2847}
(\byear{2013})
\end{barticle}
\endbibitem

\bibitem[\protect\citeauthoryear{Schuster et~al.}{2023}]{schuster2023spatial}
\begin{barticle}
\bauthor{\bsnm{Schuster}, \binits{J.}},
\bauthor{\bsnm{Mittermayer}, \binits{M.}},
\bauthor{\bsnm{Maidl}, \binits{F.-X.}},
\bauthor{\bsnm{N{\"a}tscher}, \binits{L.}},
\bauthor{\bsnm{H{\"u}lsbergen}, \binits{K.-J.}}:
\batitle{Spatial variability of soil properties, nitrogen balance and nitrate
  leaching using digital methods on heterogeneous arable fields in southern
  germany}.
\bjtitle{Precision Agriculture}
\bvolume{24}(\bissue{2}),
\bfpage{647}--\blpage{676}
(\byear{2023})
\end{barticle}
\endbibitem

\bibitem[\protect\citeauthoryear{Dewan et~al.}{2023}]{dewan2023development}
\begin{barticle}
\bauthor{\bsnm{Dewan}, \binits{G.}},
\bauthor{\bsnm{Singh}, \binits{M.}},
\bauthor{\bsnm{Sharma}, \binits{A.}}:
\batitle{Development of algorithm for variable nitrogen rate application in
  cotton crop by implementing tractor mounted nitrogen sensor}.
\bjtitle{Journal of Plant Nutrition}
\bvolume{46}(\bissue{20}),
\bfpage{4735}--\blpage{4753}
(\byear{2023})
\end{barticle}
\endbibitem

\bibitem[\protect\citeauthoryear{Stachniss et~al.}{2005}]{stachniss2005rss}
\begin{bchapter}
\bauthor{\bsnm{Stachniss}, \binits{C.}},
\bauthor{\bsnm{Grisetti}, \binits{G.}},
\bauthor{\bsnm{Burgard}, \binits{W.}}:
\bctitle{Information gain-based exploration using rao-blackwellized particle
  filters}.
In: \bbtitle{Proc. of Robotics: Science and Systems (RSS)},
pp. \bfpage{65}--\blpage{72}
(\byear{2005}).
\burl{https://www.ipb.uni-bonn.de/wp-content/papercite-data/pdf/stachniss05rss.pdf}
\end{bchapter}
\endbibitem

\bibitem[\protect\citeauthoryear{Morrison et~al.}{2019}]{morrison2019multi}
\begin{bchapter}
\bauthor{\bsnm{Morrison}, \binits{D.}},
\bauthor{\bsnm{Corke}, \binits{P.}},
\bauthor{\bsnm{Leitner}, \binits{J.}}:
\bctitle{Multi-view picking: Next-best-view reaching for improved grasping in
  clutter}.
In: \bbtitle{Proc. of IEEE International Conference on Robotics and Automation
  (ICRA)},
pp. \bfpage{8762}--\blpage{8768}
(\byear{2019})
\end{bchapter}
\endbibitem

\bibitem[\protect\citeauthoryear{Rehman and Miura}{2021}]{rehman2021viewpoint}
\begin{bchapter}
\bauthor{\bsnm{Rehman}, \binits{H.U.}},
\bauthor{\bsnm{Miura}, \binits{J.}}:
\bctitle{Viewpoint planning for automated fruit harvesting using deep
  learning}.
In: \bbtitle{Proc. of IEEE/SICE International Symposium on System Integration
  (SII)},
pp. \bfpage{409}--\blpage{414}
(\byear{2021})
\end{bchapter}
\endbibitem

\bibitem[\protect\citeauthoryear{Stache et~al.}{2023}]{stache2021adaptive}
\begin{bchapter}
\bauthor{\bsnm{Stache}, \binits{F.}},
\bauthor{\bsnm{Westheider}, \binits{J.}},
\bauthor{\bsnm{Magistri}, \binits{F.}},
\bauthor{\bsnm{Stachniss}, \binits{C.}},
\bauthor{\bsnm{Popović}, \binits{M.}}:
\bctitle{Adaptive path planning for uavs for multi-resolution semantic
  segmentation},
vol. \bseriesno{159},
p. \bfpage{104288}
(\byear{2023}).
\doiurl{10.1016/j.robot.2022.104288} .
\burl{https://www.sciencedirect.com/science/article/pii/S0921889022001774}
\end{bchapter}
\endbibitem

\bibitem[\protect\citeauthoryear{Sun et~al.}{2023}]{sun2023object}
\begin{barticle}
\bauthor{\bsnm{Sun}, \binits{T.}},
\bauthor{\bsnm{Zhang}, \binits{W.}},
\bauthor{\bsnm{Miao}, \binits{Z.}},
\bauthor{\bsnm{Zhang}, \binits{Z.}},
\bauthor{\bsnm{Li}, \binits{N.}}:
\batitle{Object localization methodology in occluded agricultural environments
  through deep learning and active sensing}.
\bjtitle{Computers and Electronics in Agriculture}
\bvolume{212},
\bfpage{108141}
(\byear{2023})
\end{barticle}
\endbibitem

\bibitem[\protect\citeauthoryear{Foix et~al.}{2018}]{foix2018task}
\begin{barticle}
\bauthor{\bsnm{Foix}, \binits{S.}},
\bauthor{\bsnm{Aleny{\`a}}, \binits{G.}},
\bauthor{\bsnm{Torras}, \binits{C.}}:
\batitle{Task-driven active sensing framework applied to leaf probing}.
\bjtitle{Computers and Electronics in Agriculture}
\bvolume{147},
\bfpage{166}--\blpage{175}
(\byear{2018})
\end{barticle}
\endbibitem

\bibitem[\protect\citeauthoryear{Wang et~al.}{2022}]{wang2022unsupervised}
\begin{bchapter}
\bauthor{\bsnm{Wang}, \binits{X.}},
\bauthor{\bsnm{Lian}, \binits{L.}},
\bauthor{\bsnm{Yu}, \binits{S.X.}}:
\bctitle{Unsupervised selective labeling for more effective semi-supervised
  learning}.
In: \bbtitle{Proc. of the European Conference on Computer Vision},
pp. \bfpage{427}--\blpage{445}
(\byear{2022}).
\bcomment{Springer}
\end{bchapter}
\endbibitem

\bibitem[\protect\citeauthoryear{Gawlikowski et~al.}{2023}]{Gawlikowski2023}
\begin{bbook}
\bauthor{\bsnm{Gawlikowski}, \binits{J.}},
\bauthor{\bsnm{Tassi}, \binits{C.R.N.}},
\bauthor{\bsnm{Ali}, \binits{M.}},
\bauthor{\bsnm{Lee}, \binits{J.}},
\bauthor{\bsnm{Humt}, \binits{M.}},
\bauthor{\bsnm{Feng}, \binits{J.}},
\bauthor{\bsnm{Kruspe}, \binits{A.}},
\bauthor{\bsnm{Triebel}, \binits{R.}},
\bauthor{\bsnm{Jung}, \binits{P.}},
\bauthor{\bsnm{Roscher}, \binits{R.}},
\bauthor{\bsnm{Shahzad}, \binits{M.}},
\bauthor{\bsnm{Yang}, \binits{W.}},
\bauthor{\bsnm{Bamler}, \binits{R.}},
\bauthor{\bsnm{Zhu}, \binits{X.X.}}:
\bbtitle{{A Survey of Uncertainty in Deep Neural Networks}}
vol. \bseriesno{0123456789}.
\bpublisher{Springer}, \blocation{???}
(\byear{2023}).
\doiurl{10.1007/s10462-023-10562-9} .
\burl{https://doi.org/10.1007/s10462-023-10562-9}
\end{bbook}
\endbibitem

\bibitem[\protect\citeauthoryear{Roscher et~al.}{2020}]{roscher2020explainable}
\begin{barticle}
\bauthor{\bsnm{Roscher}, \binits{R.}},
\bauthor{\bsnm{Bohn}, \binits{B.}},
\bauthor{\bsnm{Duarte}, \binits{M.F.}},
\bauthor{\bsnm{Garcke}, \binits{J.}}:
\batitle{Explainable machine learning for scientific insights and discoveries}.
\bjtitle{{IEEE} Access}
\bvolume{8},
\bfpage{42200}--\blpage{42216}
(\byear{2020})
\end{barticle}
\endbibitem

\bibitem[\protect\citeauthoryear{Tseng et~al.}{2022}]{tseng2021timl}
\begin{bchapter}
\bauthor{\bsnm{Tseng}, \binits{G.}},
\bauthor{\bsnm{Kerner}, \binits{H.}},
\bauthor{\bsnm{Rolnick}, \binits{D.}}:
\bctitle{{TIML}: Task-informed meta-learning for crop type mapping}.
In: \bbtitle{Proc. of AAAI Workshop on AI for Agriculture and Food Systems}
(\byear{2022}).
\burl{https://openreview.net/forum?id=de0KufElojN}
\end{bchapter}
\endbibitem

\bibitem[\protect\citeauthoryear{Chen et~al.}{2022}]{chen2022performance}
\begin{barticle}
\bauthor{\bsnm{Chen}, \binits{D.}},
\bauthor{\bsnm{Lu}, \binits{Y.}},
\bauthor{\bsnm{Li}, \binits{Z.}},
\bauthor{\bsnm{Young}, \binits{S.}}:
\batitle{Performance evaluation of deep transfer learning on multi-class
  identification of common weed species in cotton production systems}.
\bjtitle{Computers and Electronics in Agriculture}
\bvolume{198},
\bfpage{107091}
(\byear{2022})
\end{barticle}
\endbibitem

\bibitem[\protect\citeauthoryear{Whang et~al.}{2023}]{Whang2023}
\begin{barticle}
\bauthor{\bsnm{Whang}, \binits{S.E.}},
\bauthor{\bsnm{Roh}, \binits{Y.}},
\bauthor{\bsnm{Song}, \binits{H.}},
\bauthor{\bsnm{Lee}, \binits{J.G.}}:
\batitle{{Data collection and quality challenges in deep learning: a
  data-centric AI perspective}}.
\bjtitle{VLDB Journal}
(\byear{2023})
\doiurl{10.1007/s00778-022-00775-9}
{\href{https://arxiv.org/abs/2112.06409}{{arXiv:2112.06409}}}
\end{barticle}
\endbibitem

\bibitem[\protect\citeauthoryear{Kerry et~al.}{2010}]{kerry2010sampling}
\begin{botherref}
\oauthor{\bsnm{Kerry}, \binits{R.}},
\oauthor{\bsnm{Oliver}, \binits{M.}},
\oauthor{\bsnm{Frogbrook}, \binits{Z.}}:
Sampling in precision agriculture.
Geostatistical applications for precision agriculture,
35--63
(2010)
\end{botherref}
\endbibitem

\bibitem[\protect\citeauthoryear{Luo et~al.}{2019}]{luo2019distributed}
\begin{bchapter}
\bauthor{\bsnm{Luo}, \binits{W.}},
\bauthor{\bsnm{Nam}, \binits{C.}},
\bauthor{\bsnm{Kantor}, \binits{G.}},
\bauthor{\bsnm{Sycara}, \binits{K.}}:
\bctitle{Distributed environmental modeling and adaptive sampling for
  multi-robot sensor coverage}.
In: \bbtitle{Proc. of the International Conference on Autonomous Agents and
  MultiAgent Systems},
pp. \bfpage{1488}--\blpage{1496}
(\byear{2019})
\end{bchapter}
\endbibitem

\bibitem[\protect\citeauthoryear{Abbas et~al.}{2021}]{abbas2021tomato}
\begin{barticle}
\bauthor{\bsnm{Abbas}, \binits{A.}},
\bauthor{\bsnm{Jain}, \binits{S.}},
\bauthor{\bsnm{Gour}, \binits{M.}},
\bauthor{\bsnm{Vankudothu}, \binits{S.}}:
\batitle{Tomato plant disease detection using transfer learning with c-gan
  synthetic images}.
\bjtitle{Computers and Electronics in Agriculture}
\bvolume{187},
\bfpage{106279}
(\byear{2021})
\end{barticle}
\endbibitem

\bibitem[\protect\citeauthoryear{Zhang et~al.}{2020}]{zhang2020overview}
\begin{barticle}
\bauthor{\bsnm{Zhang}, \binits{X.}},
\bauthor{\bsnm{Cao}, \binits{Z.}},
\bauthor{\bsnm{Dong}, \binits{W.}}:
\batitle{Overview of edge computing in the agricultural internet of things: key
  technologies, applications, challenges}.
\bjtitle{IEEE Access}
\bvolume{8},
\bfpage{141748}--\blpage{141761}
(\byear{2020})
\end{barticle}
\endbibitem

\bibitem[\protect\citeauthoryear{Koedel et~al.}{2022}]{koedel2022challenges}
\begin{barticle}
\bauthor{\bsnm{Koedel}, \binits{U.}},
\bauthor{\bsnm{Schuetze}, \binits{C.}},
\bauthor{\bsnm{Fischer}, \binits{P.}},
\bauthor{\bsnm{Bussmann}, \binits{I.}},
\bauthor{\bsnm{Sauer}, \binits{P.K.}},
\bauthor{\bsnm{Nixdorf}, \binits{E.}},
\bauthor{\bsnm{Kalbacher}, \binits{T.}},
\bauthor{\bsnm{Wichert}, \binits{V.}},
\bauthor{\bsnm{Rechid}, \binits{D.}},
\bauthor{\bsnm{Bouwer}, \binits{L.M.}}, \betal:
\batitle{Challenges in the evaluation of observational data trustworthiness
  from a data producers viewpoint (fair+)}.
\bjtitle{Frontiers in Environmental Science}
\bvolume{9},
\bfpage{772666}
(\byear{2022})
\end{barticle}
\endbibitem

\bibitem[\protect\citeauthoryear{Li et~al.}{2023}]{li2023foundation}
\begin{botherref}
\oauthor{\bsnm{Li}, \binits{J.}},
\oauthor{\bsnm{Xu}, \binits{M.}},
\oauthor{\bsnm{Xiang}, \binits{L.}},
\oauthor{\bsnm{Chen}, \binits{D.}},
\oauthor{\bsnm{Zhuang}, \binits{W.}},
\oauthor{\bsnm{Yin}, \binits{X.}},
\oauthor{\bsnm{Li}, \binits{Z.}}:
Foundation models in smart agriculture: Basics, opportunities, and challenges.
arXiv preprint arXiv:2308.06668
(2023)
\end{botherref}
\endbibitem

\bibitem[\protect\citeauthoryear{Autz et~al.}{2022}]{autz2022pitfalls}
\begin{botherref}
\oauthor{\bsnm{Autz}, \binits{J.}},
\oauthor{\bsnm{Mishra}, \binits{S.K.}},
\oauthor{\bsnm{Herrmann}, \binits{L.}},
\oauthor{\bsnm{Hertzberg}, \binits{J.}}:
The pitfalls of transfer learning in computer vision for agriculture.
42. GIL-Jahrestagung, K{\"u}nstliche Intelligenz in der Agrar-und
  Ern{\"a}hrungswirtschaft
(2022)
\end{botherref}
\endbibitem

\bibitem[\protect\citeauthoryear{Yang et~al.}{2020}]{yang2020transfer}
\begin{bbook}
\bauthor{\bsnm{Yang}, \binits{Q.}},
\bauthor{\bsnm{Zhang}, \binits{Y.}},
\bauthor{\bsnm{Dai}, \binits{W.}},
\bauthor{\bsnm{Pan}, \binits{S.J.}}:
\bbtitle{Transfer Learning},
p. \bfpage{380}.
\bpublisher{Cambridge University Press}, \blocation{???}
(\byear{2020}).
\doiurl{10.1017/9781139061773}
\end{bbook}
\endbibitem

\bibitem[\protect\citeauthoryear{Espejo-Garcia
  et~al.}{2020}]{espejo2020towards}
\begin{barticle}
\bauthor{\bsnm{Espejo-Garcia}, \binits{B.}},
\bauthor{\bsnm{Mylonas}, \binits{N.}},
\bauthor{\bsnm{Athanasakos}, \binits{L.}},
\bauthor{\bsnm{Fountas}, \binits{S.}},
\bauthor{\bsnm{Vasilakoglou}, \binits{I.}}:
\batitle{Towards weeds identification assistance through transfer learning}.
\bjtitle{Computers and Electronics in Agriculture}
\bvolume{171},
\bfpage{105306}
(\byear{2020})
\end{barticle}
\endbibitem

\bibitem[\protect\citeauthoryear{Al~Sahili and Awad}{2022}]{al2022power}
\begin{barticle}
\bauthor{\bsnm{Al~Sahili}, \binits{Z.}},
\bauthor{\bsnm{Awad}, \binits{M.}}:
\batitle{The power of transfer learning in agricultural applications: Agrinet}.
\bjtitle{Frontiers in Plant Science}
\bvolume{13},
\bfpage{992700}
(\byear{2022})
\end{barticle}
\endbibitem

\bibitem[\protect\citeauthoryear{Magistri et~al.}{2023}]{magistri2023one}
\begin{barticle}
\bauthor{\bsnm{Magistri}, \binits{F.}},
\bauthor{\bsnm{Weyler}, \binits{J.}},
\bauthor{\bsnm{Gogoll}, \binits{D.}},
\bauthor{\bsnm{Lottes}, \binits{P.}},
\bauthor{\bsnm{Behley}, \binits{J.}},
\bauthor{\bsnm{Petrinic}, \binits{N.}},
\bauthor{\bsnm{Stachniss}, \binits{C.}}:
\batitle{From one field to another—unsupervised domain adaptation for
  semantic segmentation in agricultural robotics}.
\bjtitle{Computers and Electronics in Agriculture}
\bvolume{212},
\bfpage{108114}
(\byear{2023})
\end{barticle}
\endbibitem

\bibitem[\protect\citeauthoryear{Wan
  et~al.}{2023}]{doi:10.34133/icomputing.0058}
\begin{barticle}
\bauthor{\bsnm{Wan}, \binits{T.}},
\bauthor{\bsnm{Xu}, \binits{K.}},
\bauthor{\bsnm{Yu}, \binits{T.}},
\bauthor{\bsnm{Wang}, \binits{X.}},
\bauthor{\bsnm{Feng}, \binits{D.}},
\bauthor{\bsnm{Ding}, \binits{B.}},
\bauthor{\bsnm{Wang}, \binits{H.}}:
\batitle{{A Survey of Deep Active Learning for Foundation Models}}.
\bjtitle{Intelligent Computing}
(\byear{2023})
\doiurl{10.34133/icomputing.0058}
\end{barticle}
\endbibitem

\bibitem[\protect\citeauthoryear{Tifrea et~al.}{2022}]{tifrea2022uniform}
\begin{botherref}
\oauthor{\bsnm{Tifrea}, \binits{A.}},
\oauthor{\bsnm{Clarysse}, \binits{J.}},
\oauthor{\bsnm{Yang}, \binits{F.}}:
Uniform versus uncertainty sampling: When being active is less efficient than
  staying passive.
arXiv preprint arXiv:2212.00772
(2022)
\end{botherref}
\endbibitem

\bibitem[\protect\citeauthoryear{Thenmozhi and Reddy}{2019}]{thenmozhi2019crop}
\begin{barticle}
\bauthor{\bsnm{Thenmozhi}, \binits{K.}},
\bauthor{\bsnm{Reddy}, \binits{U.S.}}:
\batitle{Crop pest classification based on deep convolutional neural network
  and transfer learning}.
\bjtitle{Computers and Electronics in Agriculture}
\bvolume{164},
\bfpage{104906}
(\byear{2019})
\end{barticle}
\endbibitem

\bibitem[\protect\citeauthoryear{Chen et~al.}{2020}]{chen2020using}
\begin{barticle}
\bauthor{\bsnm{Chen}, \binits{J.}},
\bauthor{\bsnm{Chen}, \binits{J.}},
\bauthor{\bsnm{Zhang}, \binits{D.}},
\bauthor{\bsnm{Sun}, \binits{Y.}},
\bauthor{\bsnm{Nanehkaran}, \binits{Y.A.}}:
\batitle{Using deep transfer learning for image-based plant disease
  identification}.
\bjtitle{Computers and Electronics in Agriculture}
\bvolume{173},
\bfpage{105393}
(\byear{2020})
\end{barticle}
\endbibitem

\bibitem[\protect\citeauthoryear{Dara et~al.}{2022}]{dara2022recommendations}
\begin{barticle}
\bauthor{\bsnm{Dara}, \binits{R.}},
\bauthor{\bsnm{Hazrati~Fard}, \binits{S.M.}},
\bauthor{\bsnm{Kaur}, \binits{J.}}:
\batitle{Recommendations for ethical and responsible use of artificial
  intelligence in digital agriculture}.
\bjtitle{Frontiers in Artificial Intelligence}
\bvolume{5},
\bfpage{884192}
(\byear{2022})
\end{barticle}
\endbibitem

\bibitem[\protect\citeauthoryear{Liang et~al.}{2022}]{Liang2022}
\begin{barticle}
\bauthor{\bsnm{Liang}, \binits{W.}},
\bauthor{\bsnm{Tadesse}, \binits{G.A.}},
\bauthor{\bsnm{Ho}, \binits{D.}},
\bauthor{\bsnm{Li}, \binits{F.F.}},
\bauthor{\bsnm{Zaharia}, \binits{M.}},
\bauthor{\bsnm{Zhang}, \binits{C.}},
\bauthor{\bsnm{Zou}, \binits{J.}}:
\batitle{{Advances, challenges and opportunities in creating data for
  trustworthy AI}}.
\bjtitle{Nature Machine Intelligence}
\bvolume{4}(\bissue{8}),
\bfpage{669}--\blpage{677}
(\byear{2022})
\doiurl{10.1038/s42256-022-00516-1}
\end{barticle}
\endbibitem

\bibitem[\protect\citeauthoryear{Kumar et~al.}{2021}]{kumar2021sp2f}
\begin{barticle}
\bauthor{\bsnm{Kumar}, \binits{R.}},
\bauthor{\bsnm{Kumar}, \binits{P.}},
\bauthor{\bsnm{Tripathi}, \binits{R.}},
\bauthor{\bsnm{Gupta}, \binits{G.P.}},
\bauthor{\bsnm{Gadekallu}, \binits{T.R.}},
\bauthor{\bsnm{Srivastava}, \binits{G.}}:
\batitle{Sp2f: A secured privacy-preserving framework for smart agricultural
  unmanned aerial vehicles}.
\bjtitle{Computer Networks}
\bvolume{187},
\bfpage{107819}
(\byear{2021})
\end{barticle}
\endbibitem

\bibitem[\protect\citeauthoryear{Gardezi et~al.}{2023}]{gardezi2023artificial}
\begin{botherref}
\oauthor{\bsnm{Gardezi}, \binits{M.}},
\oauthor{\bsnm{Joshi}, \binits{B.}},
\oauthor{\bsnm{Rizzo}, \binits{D.M.}},
\oauthor{\bsnm{Ryan}, \binits{M.}},
\oauthor{\bsnm{Prutzer}, \binits{E.}},
\oauthor{\bsnm{Brugler}, \binits{S.}},
\oauthor{\bsnm{Dadkhah}, \binits{A.}}:
Artificial intelligence in farming: Challenges and opportunities for building
  trust.
Agronomy Journal
(2023)
\end{botherref}
\endbibitem

\bibitem[\protect\citeauthoryear{Aggarwal et~al.}{2023}]{aggarwal2023federated}
\begin{barticle}
\bauthor{\bsnm{Aggarwal}, \binits{M.}},
\bauthor{\bsnm{Khullar}, \binits{V.}},
\bauthor{\bsnm{Goyal}, \binits{N.}},
\bauthor{\bsnm{Gautam}, \binits{R.}},
\bauthor{\bsnm{Alblehai}, \binits{F.}},
\bauthor{\bsnm{Elghatwary}, \binits{M.}},
\bauthor{\bsnm{Singh}, \binits{A.}}:
\batitle{Federated transfer learning for rice-leaf disease classification
  across multiclient cross-silo datasets}.
\bjtitle{Agronomy}
\bvolume{13}(\bissue{10}),
\bfpage{2483}
(\byear{2023})
\end{barticle}
\endbibitem

\bibitem[\protect\citeauthoryear{}{2020}]{nations2020sustainable}
\begin{botherref}
United Nations: The Sustainable Development Goals Report
(2020)
\end{botherref}
\endbibitem

\bibitem[\protect\citeauthoryear{Khanna et~al.}{2022}]{khanna2022digital}
\begin{barticle}
\bauthor{\bsnm{Khanna}, \binits{M.}},
\bauthor{\bsnm{Atallah}, \binits{S.S.}},
\bauthor{\bsnm{Kar}, \binits{S.}},
\bauthor{\bsnm{Sharma}, \binits{B.}},
\bauthor{\bsnm{Wu}, \binits{L.}},
\bauthor{\bsnm{Yu}, \binits{C.}},
\bauthor{\bsnm{Chowdhary}, \binits{G.}},
\bauthor{\bsnm{Soman}, \binits{C.}},
\bauthor{\bsnm{Guan}, \binits{K.}}:
\batitle{Digital transformation for a sustainable agriculture in the united
  states: Opportunities and challenges}.
\bjtitle{Agricultural Economics}
\bvolume{53}(\bissue{6}),
\bfpage{924}--\blpage{937}
(\byear{2022})
\end{barticle}
\endbibitem

\bibitem[\protect\citeauthoryear{Van~Wynsberghe}{2021}]{van2021sustainable}
\begin{barticle}
\bauthor{\bsnm{Van~Wynsberghe}, \binits{A.}}:
\batitle{Sustainable ai: Ai for sustainability and the sustainability of ai}.
\bjtitle{AI and Ethics}
\bvolume{1}(\bissue{3}),
\bfpage{213}--\blpage{218}
(\byear{2021})
\end{barticle}
\endbibitem

\bibitem[\protect\citeauthoryear{Sanchez-Iborra and
  Skarmeta}{2020}]{sanchez2020tinyml}
\begin{barticle}
\bauthor{\bsnm{Sanchez-Iborra}, \binits{R.}},
\bauthor{\bsnm{Skarmeta}, \binits{A.F.}}:
\batitle{Tinyml-enabled frugal smart objects: Challenges and opportunities}.
\bjtitle{IEEE Circuits and Systems Magazine}
\bvolume{20}(\bissue{3}),
\bfpage{4}--\blpage{18}
(\byear{2020})
\end{barticle}
\endbibitem

\bibitem[\protect\citeauthoryear{Al-Jarrah et~al.}{2015}]{al2015efficient}
\begin{barticle}
\bauthor{\bsnm{Al-Jarrah}, \binits{O.Y.}},
\bauthor{\bsnm{Yoo}, \binits{P.D.}},
\bauthor{\bsnm{Muhaidat}, \binits{S.}},
\bauthor{\bsnm{Karagiannidis}, \binits{G.K.}},
\bauthor{\bsnm{Taha}, \binits{K.}}:
\batitle{Efficient machine learning for big data: A review}.
\bjtitle{Big Data Research}
\bvolume{2}(\bissue{3}),
\bfpage{87}--\blpage{93}
(\byear{2015})
\end{barticle}
\endbibitem

\bibitem[\protect\citeauthoryear{McCool et~al.}{2017}]{mccool2017mixtures}
\begin{barticle}
\bauthor{\bsnm{McCool}, \binits{C.}},
\bauthor{\bsnm{Perez}, \binits{T.}},
\bauthor{\bsnm{Upcroft}, \binits{B.}}:
\batitle{Mixtures of lightweight deep convolutional neural networks: Applied to
  agricultural robotics}.
\bjtitle{IEEE Robotics and Automation Letters}
\bvolume{2}(\bissue{3}),
\bfpage{1344}--\blpage{1351}
(\byear{2017})
\end{barticle}
\endbibitem

\bibitem[\protect\citeauthoryear{Li et~al.}{2019}]{li2019real}
\begin{barticle}
\bauthor{\bsnm{Li}, \binits{N.}},
\bauthor{\bsnm{Zhang}, \binits{X.}},
\bauthor{\bsnm{Zhang}, \binits{C.}},
\bauthor{\bsnm{Guo}, \binits{H.}},
\bauthor{\bsnm{Sun}, \binits{Z.}},
\bauthor{\bsnm{Wu}, \binits{X.}}:
\batitle{Real-time crop recognition in transplanted fields with prominent weed
  growth: a visual-attention-based approach}.
\bjtitle{IEEE Access}
\bvolume{7},
\bfpage{185310}--\blpage{185321}
(\byear{2019})
\end{barticle}
\endbibitem

\bibitem[\protect\citeauthoryear{Paudel et~al.}{2021}]{paudel2021machine}
\begin{barticle}
\bauthor{\bsnm{Paudel}, \binits{D.}},
\bauthor{\bsnm{Boogaard}, \binits{H.}},
\bauthor{\bsnm{Wit}, \binits{A.}},
\bauthor{\bsnm{Janssen}, \binits{S.}},
\bauthor{\bsnm{Osinga}, \binits{S.}},
\bauthor{\bsnm{Pylianidis}, \binits{C.}},
\bauthor{\bsnm{Athanasiadis}, \binits{I.N.}}:
\batitle{Machine learning for large-scale crop yield forecasting}.
\bjtitle{Agricultural Systems}
\bvolume{187},
\bfpage{103016}
(\byear{2021})
\end{barticle}
\endbibitem

\bibitem[\protect\citeauthoryear{Condran et~al.}{2022}]{condran2022machine}
\begin{barticle}
\bauthor{\bsnm{Condran}, \binits{S.}},
\bauthor{\bsnm{Bewong}, \binits{M.}},
\bauthor{\bsnm{Islam}, \binits{M.Z.}},
\bauthor{\bsnm{Maphosa}, \binits{L.}},
\bauthor{\bsnm{Zheng}, \binits{L.}}:
\batitle{Machine learning in precision agriculture: a survey on trends,
  applications and evaluations over two decades}.
\bjtitle{IEEE Access}
\bvolume{10},
\bfpage{73786}--\blpage{73803}
(\byear{2022})
\end{barticle}
\endbibitem

\bibitem[\protect\citeauthoryear{A.~Ramezan et~al.}{2019}]{a2019evaluation}
\begin{barticle}
\bauthor{\bsnm{A.~Ramezan}, \binits{C.}},
\bauthor{\bsnm{A.~Warner}, \binits{T.}},
\bauthor{\bsnm{E.~Maxwell}, \binits{A.}}:
\batitle{Evaluation of sampling and cross-validation tuning strategies for
  regional-scale machine learning classification}.
\bjtitle{Remote Sensing}
\bvolume{11}(\bissue{2}),
\bfpage{185}
(\byear{2019})
\end{barticle}
\endbibitem

\bibitem[\protect\citeauthoryear{Du et~al.}{2022}]{du2022estimating}
\begin{barticle}
\bauthor{\bsnm{Du}, \binits{L.}},
\bauthor{\bsnm{Yang}, \binits{H.}},
\bauthor{\bsnm{Song}, \binits{X.}},
\bauthor{\bsnm{Wei}, \binits{N.}},
\bauthor{\bsnm{Yu}, \binits{C.}},
\bauthor{\bsnm{Wang}, \binits{W.}},
\bauthor{\bsnm{Zhao}, \binits{Y.}}:
\batitle{Estimating leaf area index of maize using uav-based digital imagery
  and machine learning methods}.
\bjtitle{Scientific Reports}
\bvolume{12}(\bissue{1}),
\bfpage{15937}
(\byear{2022})
\end{barticle}
\endbibitem

\bibitem[\protect\citeauthoryear{Chung et~al.}{2019}]{chung2019slice}
\begin{bchapter}
\bauthor{\bsnm{Chung}, \binits{Y.}},
\bauthor{\bsnm{Kraska}, \binits{T.}},
\bauthor{\bsnm{Polyzotis}, \binits{N.}},
\bauthor{\bsnm{Tae}, \binits{K.H.}},
\bauthor{\bsnm{Whang}, \binits{S.E.}}:
\bctitle{Slice finder: Automated data slicing for model validation}.
In: \bbtitle{Proc. of the International Conference on Data Engineering (ICDE)},
pp. \bfpage{1550}--\blpage{1553}
(\byear{2019})
\end{bchapter}
\endbibitem

\bibitem[\protect\citeauthoryear{Koh et~al.}{2021}]{koh2021wilds}
\begin{bchapter}
\bauthor{\bsnm{Koh}, \binits{P.W.}},
\bauthor{\bsnm{Sagawa}, \binits{S.}},
\bauthor{\bsnm{Marklund}, \binits{H.}},
\bauthor{\bsnm{Xie}, \binits{S.M.}},
\bauthor{\bsnm{Zhang}, \binits{M.}},
\bauthor{\bsnm{Balsubramani}, \binits{A.}},
\bauthor{\bsnm{Hu}, \binits{W.}},
\bauthor{\bsnm{Yasunaga}, \binits{M.}},
\bauthor{\bsnm{Phillips}, \binits{R.L.}},
\bauthor{\bsnm{Gao}, \binits{I.}}, \betal:
\bctitle{Wilds: A benchmark of in-the-wild distribution shifts}.
In: \bbtitle{Proc. of the International Conference on Machine Learning},
pp. \bfpage{5637}--\blpage{5664}
(\byear{2021})
\end{bchapter}
\endbibitem

\bibitem[\protect\citeauthoryear{Meyer and Pebesma}{2021}]{meyer2021predicting}
\begin{barticle}
\bauthor{\bsnm{Meyer}, \binits{H.}},
\bauthor{\bsnm{Pebesma}, \binits{E.}}:
\batitle{Predicting into unknown space? estimating the area of applicability of
  spatial prediction models}.
\bjtitle{Methods in Ecology and Evolution}
\bvolume{12}(\bissue{9}),
\bfpage{1620}--\blpage{1633}
(\byear{2021})
\end{barticle}
\endbibitem

\bibitem[\protect\citeauthoryear{Schramowski et~al.}{2020}]{Schramowski2020a}
\begin{barticle}
\bauthor{\bsnm{Schramowski}, \binits{P.}},
\bauthor{\bsnm{Stammer}, \binits{W.}},
\bauthor{\bsnm{Teso}, \binits{S.}},
\bauthor{\bsnm{Brugger}, \binits{A.}},
\bauthor{\bsnm{Herbert}, \binits{F.}},
\bauthor{\bsnm{Shao}, \binits{X.}},
\bauthor{\bsnm{Luigs}, \binits{H.G.}},
\bauthor{\bsnm{Mahlein}, \binits{A.K.}},
\bauthor{\bsnm{Kersting}, \binits{K.}}:
\batitle{{Making deep neural networks right for the right scientific reasons by
  interacting with their explanations}}.
\bjtitle{Nature Machine Intelligence}
\bvolume{2}(\bissue{8}),
\bfpage{476}--\blpage{486}
(\byear{2020})
\doiurl{10.1038/s42256-020-0212-3}
{\href{https://arxiv.org/abs/2001.05371}{{arXiv:2001.05371}}}
\end{barticle}
\endbibitem

\bibitem[\protect\citeauthoryear{Kierdorf and
  Roscher}{2023}]{kierdorf2023reliability}
\begin{barticle}
\bauthor{\bsnm{Kierdorf}, \binits{J.}},
\bauthor{\bsnm{Roscher}, \binits{R.}}:
\batitle{Reliability scores from saliency map clusters for improved image-based
  harvest-readiness prediction in cauliflower}.
\bjtitle{IEEE Geoscience and Remote Sensing Letters}
(\byear{2023})
\doiurl{10.1109/LGRS.2023.3293802}
\end{barticle}
\endbibitem

\bibitem[\protect\citeauthoryear{Lu et~al.}{2022}]{lu2022generative}
\begin{barticle}
\bauthor{\bsnm{Lu}, \binits{Y.}},
\bauthor{\bsnm{Chen}, \binits{D.}},
\bauthor{\bsnm{Olaniyi}, \binits{E.}},
\bauthor{\bsnm{Huang}, \binits{Y.}}:
\batitle{Generative adversarial networks (gans) for image augmentation in
  agriculture: A systematic review}.
\bjtitle{Computers and Electronics in Agriculture}
\bvolume{200},
\bfpage{107208}
(\byear{2022})
\end{barticle}
\endbibitem

\bibitem[\protect\citeauthoryear{Roscher et~al.}{2020}]{roscher2020explain}
\begin{barticle}
\bauthor{\bsnm{Roscher}, \binits{R.}},
\bauthor{\bsnm{Bohn}, \binits{B.}},
\bauthor{\bsnm{Duarte}, \binits{M.}},
\bauthor{\bsnm{Garcke}, \binits{J.}}:
\batitle{Explain it to me - facing remote sensing challenges in the bio-and
  geosciences with explainable machine learning}.
\bjtitle{ISPRS Annals of Photogrammetry, Remote Sensing and Spatial Information
  Sciences}
\bvolume{5},
\bfpage{817}--\blpage{824}
(\byear{2020})
\end{barticle}
\endbibitem

\bibitem[\protect\citeauthoryear{Loucks et~al.}{2005}]{loucks2005model}
\begin{botherref}
\oauthor{\bsnm{Loucks}, \binits{D.}},
\oauthor{\bsnm{Van~Beek}, \binits{E.}},
\oauthor{\bsnm{Stedinger}, \binits{J.}},
\oauthor{\bsnm{Dijkman}, \binits{J.}},
\oauthor{\bsnm{Villars}, \binits{M.}}:
Model sensitivity and uncertainty analysis.
Water resources systems planning and management,
255--290
(2005)
\end{botherref}
\endbibitem

\bibitem[\protect\citeauthoryear{Visser et~al.}{2021}]{visser2021imprecision}
\begin{barticle}
\bauthor{\bsnm{Visser}, \binits{O.}},
\bauthor{\bsnm{Sippel}, \binits{S.R.}},
\bauthor{\bsnm{Thiemann}, \binits{L.}}:
\batitle{Imprecision farming? examining the (in) accuracy and risks of digital
  agriculture}.
\bjtitle{Journal of Rural Studies}
\bvolume{86},
\bfpage{623}--\blpage{632}
(\byear{2021})
\end{barticle}
\endbibitem

\bibitem[\protect\citeauthoryear{Guo et~al.}{2017}]{guo2017calibration}
\begin{bchapter}
\bauthor{\bsnm{Guo}, \binits{C.}},
\bauthor{\bsnm{Pleiss}, \binits{G.}},
\bauthor{\bsnm{Sun}, \binits{Y.}},
\bauthor{\bsnm{Weinberger}, \binits{K.Q.}}:
\bctitle{On calibration of modern neural networks}.
In: \bbtitle{Proc. of the International Conference on Machine Learning},
pp. \bfpage{1321}--\blpage{1330}
(\byear{2017})
\end{bchapter}
\endbibitem

\bibitem[\protect\citeauthoryear{Wilson and
  Izmailov}{2020}]{wilson2020bayesian}
\begin{bchapter}
\bauthor{\bsnm{Wilson}, \binits{A.G.}},
\bauthor{\bsnm{Izmailov}, \binits{P.}}:
\bctitle{Bayesian deep learning and a probabilistic perspective of
  generalization}.
In: \bbtitle{Proc.~of the Conference on Neural Information Processing Systems
  (NeurIPS)},
pp. \bfpage{4697}--\blpage{4708}
(\byear{2020})
\end{bchapter}
\endbibitem

\bibitem[\protect\citeauthoryear{Lee et~al.}{2018}]{lee2018training}
\begin{bchapter}
\bauthor{\bsnm{Lee}, \binits{K.}},
\bauthor{\bsnm{Lee}, \binits{H.}},
\bauthor{\bsnm{Lee}, \binits{K.}},
\bauthor{\bsnm{Shin}, \binits{J.}}:
\bctitle{Training confidence-calibrated classifiers for detecting
  out-of-distribution samples}.
In: \bbtitle{Proc.~of the Int.~Conf.~on Learning Representations (ICLR)}
(\byear{2018})
\end{bchapter}
\endbibitem

\bibitem[\protect\citeauthoryear{Ovadia et~al.}{2019}]{ovadia2019can}
\begin{bchapter}
\bauthor{\bsnm{Ovadia}, \binits{Y.}},
\bauthor{\bsnm{Fertig}, \binits{E.}},
\bauthor{\bsnm{Ren}, \binits{J.}},
\bauthor{\bsnm{Nado}, \binits{Z.}},
\bauthor{\bsnm{Sculley}, \binits{D.}},
\bauthor{\bsnm{Nowozin}, \binits{S.}},
\bauthor{\bsnm{Dillon}, \binits{J.}},
\bauthor{\bsnm{Lakshminarayanan}, \binits{B.}},
\bauthor{\bsnm{Snoek}, \binits{J.}}:
\bctitle{Can you trust your model's uncertainty? evaluating predictive
  uncertainty under dataset shift}.
In: \bbtitle{Proc.~of the Conference on Neural Information Processing Systems
  (NeurIPS)},
vol. \bseriesno{32},
pp. \bfpage{13991}--\blpage{14002}
(\byear{2019})
\end{bchapter}
\endbibitem

\bibitem[\protect\citeauthoryear{Smith and Gal}{2018}]{smith2018understanding}
\begin{bchapter}
\bauthor{\bsnm{Smith}, \binits{L.}},
\bauthor{\bsnm{Gal}, \binits{Y.}}:
\bctitle{Understanding measures of uncertainty for adversarial example
  detection}.
In: \bbtitle{Proc.~of the Conf. on Uncertainty in Artificial Intelligence},
pp. \bfpage{560}--\blpage{569}
(\byear{2018})
\end{bchapter}
\endbibitem

\bibitem[\protect\citeauthoryear{Patel et~al.}{2021}]{patel2021manifold}
\begin{bchapter}
\bauthor{\bsnm{Patel}, \binits{K.}},
\bauthor{\bsnm{Beluch}, \binits{W.}},
\bauthor{\bsnm{Zhang}, \binits{D.}},
\bauthor{\bsnm{Pfeiffer}, \binits{M.}},
\bauthor{\bsnm{Yang}, \binits{B.}}:
\bctitle{On-manifold adversarial data augmentation improves uncertainty
  calibration}.
In: \bbtitle{Proc. of International Conference on Pattern Recognition},
pp. \bfpage{8029}--\blpage{8036}
(\byear{2021})
\end{bchapter}
\endbibitem

\bibitem[\protect\citeauthoryear{Lakshminarayanan
  et~al.}{2017}]{lakshminarayanan2017simple}
\begin{bchapter}
\bauthor{\bsnm{Lakshminarayanan}, \binits{B.}},
\bauthor{\bsnm{Pritzel}, \binits{A.}},
\bauthor{\bsnm{Blundell}, \binits{C.}}:
\bctitle{Simple and scalable predictive uncertainty estimation using deep
  ensembles}.
In: \bbtitle{Proc.~of the Conference on Neural Information Processing Systems
  (NeurIPS)},
vol. \bseriesno{30}
(\byear{2017})
\end{bchapter}
\endbibitem

\bibitem[\protect\citeauthoryear{Kendall and
  Gal}{2017}]{kendall2017uncertainties}
\begin{bchapter}
\bauthor{\bsnm{Kendall}, \binits{A.}},
\bauthor{\bsnm{Gal}, \binits{Y.}}:
\bctitle{What uncertainties do we need in bayesian deep learning for computer
  vision?}
In: \bbtitle{Proc.~of the Conference on Neural Information Processing Systems
  (NeurIPS)},
vol. \bseriesno{30},
pp. \bfpage{5574}--\blpage{5584}
(\byear{2017})
\end{bchapter}
\endbibitem

\bibitem[\protect\citeauthoryear{Bau et~al.}{2020}]{bau2020understanding}
\begin{barticle}
\bauthor{\bsnm{Bau}, \binits{D.}},
\bauthor{\bsnm{Zhu}, \binits{J.-Y.}},
\bauthor{\bsnm{Strobelt}, \binits{H.}},
\bauthor{\bsnm{Lapedriza}, \binits{A.}},
\bauthor{\bsnm{Zhou}, \binits{B.}},
\bauthor{\bsnm{Torralba}, \binits{A.}}:
\batitle{Understanding the role of individual units in a deep neural network}.
\bjtitle{Proc.~of the National Academy of Sciences}
\bvolume{117}(\bissue{48}),
\bfpage{30071}--\blpage{30078}
(\byear{2020})
\end{barticle}
\endbibitem

\bibitem[\protect\citeauthoryear{Adebayo et~al.}{2018}]{adebayo2018sanity}
\begin{bchapter}
\bauthor{\bsnm{Adebayo}, \binits{J.}},
\bauthor{\bsnm{Gilmer}, \binits{J.}},
\bauthor{\bsnm{Muelly}, \binits{M.}},
\bauthor{\bsnm{Goodfellow}, \binits{I.}},
\bauthor{\bsnm{Hardt}, \binits{M.}},
\bauthor{\bsnm{Kim}, \binits{B.}}:
\bctitle{Sanity checks for saliency maps}.
In: \bbtitle{Proc. of the International Conference on Neural Information
  Processing Systems},
pp. \bfpage{9525}--\blpage{9536}
(\byear{2018})
\end{bchapter}
\endbibitem

\bibitem[\protect\citeauthoryear{}{}]{zotero-3810}
\begin{botherref}
Bill \& Melinda Gates Foundation Open {{Access Policy}}.
\url{https://www.gatesfoundation.org/about/policies-and-resources/open-access-policy}.
accessed on 2023-11-13
\end{botherref}
\endbibitem

\bibitem[\protect\citeauthoryear{}{}]{zotero-3813}
\begin{botherref}
{{CGIAR Open Access Data Management Policy}}.
\url{https://cgspace.cgiar.org/bitstream/handle/10947/4488/Open\%20Access\%20Data\%20Management\%20Policy.pdf?sequence=1}.
accessed on 2023-11-13
\end{botherref}
\endbibitem

\bibitem[\protect\citeauthoryear{}{}]{zotero-3814}
\begin{botherref}
Wetter Und {{Klima}} - {{Deutscher Wetterdienst}} - {{Leistungen}} - {{Open
  Data}}.
\url{https://www.dwd.de/DE/leistungen/opendata/opendata.html}.
accessed on 2023-11-13
\end{botherref}
\endbibitem

\bibitem[\protect\citeauthoryear{Ammann et~al.}{2022a}]{ammann2022}
\begin{barticle}
\bauthor{\bsnm{Ammann}, \binits{J.}},
\bauthor{\bsnm{Walter}, \binits{A.}},
\bauthor{\bsnm{El~Benni}, \binits{N.}}:
\batitle{Adoption and perception of farm management information systems by
  future {{Swiss}} farm managers \textendash{} {{An}} online study}.
\bjtitle{Journal of Rural Studies}
\bvolume{89},
\bfpage{298}--\blpage{305}
(\byear{2022})
\doiurl{10.1016/j.jrurstud.2021.12.008}
\end{barticle}
\endbibitem

\bibitem[\protect\citeauthoryear{Ammann et~al.}{2022b}]{ammann2022a}
\begin{barticle}
\bauthor{\bsnm{Ammann}, \binits{J.}},
\bauthor{\bsnm{Umst{\"a}tter}, \binits{C.}},
\bauthor{\bsnm{El~Benni}, \binits{N.}}:
\batitle{The adoption of precision agriculture enabling technologies in
  {{Swiss}} outdoor vegetable production: A {{Delphi}} study}.
\bjtitle{Precision Agriculture}
\bvolume{23}(\bissue{4}),
\bfpage{1354}--\blpage{1374}
(\byear{2022})
\doiurl{10.1007/s11119-022-09889-0}
\end{barticle}
\endbibitem

\bibitem[\protect\citeauthoryear{Zhang et~al.}{2021}]{zhang2021}
\begin{barticle}
\bauthor{\bsnm{Zhang}, \binits{A.}},
\bauthor{\bsnm{Heath}, \binits{R.}},
\bauthor{\bsnm{McRobert}, \binits{K.}},
\bauthor{\bsnm{Llewellyn}, \binits{R.}},
\bauthor{\bsnm{Sanderson}, \binits{J.}},
\bauthor{\bsnm{Wiseman}, \binits{L.}},
\bauthor{\bsnm{Rainbow}, \binits{R.}}:
\batitle{Who will benefit from big data? {{Farmers}}' perspective on
  willingness to share farm data}.
\bjtitle{Journal of Rural Studies}
\bvolume{88},
\bfpage{346}--\blpage{353}
(\byear{2021})
\doiurl{10.1016/j.jrurstud.2021.08.006}
\end{barticle}
\endbibitem

\bibitem[\protect\citeauthoryear{Barnard}{1975}]{barnard1975}
\begin{barticle}
\bauthor{\bsnm{Barnard}, \binits{C.S.}}:
\batitle{Data in {{Agriculture}}: {{A Review}} with {{Special Reference}} to
  {{Farm Management Research}}, {{Policy}} and {{Advice}} in {{Britain}}*}.
\bjtitle{Journal of Agricultural Economics}
\bvolume{26}(\bissue{3}),
\bfpage{289}--\blpage{333}
(\byear{1975})
\doiurl{10.1111/j.1477-9552.1975.tb02332.x}
\end{barticle}
\endbibitem

\bibitem[\protect\citeauthoryear{Federer}{1955}]{federer1955}
\begin{bbook}
\bauthor{\bsnm{Federer}, \binits{W.}}:
\bbtitle{Experimental Design}.
\bpublisher{The MacMillan Company: New York}, \blocation{???}
(\byear{1955})
\end{bbook}
\endbibitem

\bibitem[\protect\citeauthoryear{Brown et~al.}{2020}]{brown2020}
\begin{barticle}
\bauthor{\bsnm{Brown}, \binits{D.}},
\bauthor{\bsnm{{Van den Bergh}}, \binits{I.}},
\bauthor{\bsnm{{de Bruin}}, \binits{S.}},
\bauthor{\bsnm{Machida}, \binits{L.}},
\bauthor{\bsnm{{van Etten}}, \binits{J.}}:
\batitle{Data synthesis for crop variety evaluation. {{A}} review}.
\bjtitle{Agronomy for Sustainable Development}
\bvolume{40}(\bissue{4}),
\bfpage{25}
(\byear{2020})
\doiurl{10.1007/s13593-020-00630-7}
\end{barticle}
\endbibitem

\bibitem[\protect\citeauthoryear{Piepho et~al.}{2011}]{piepho2011}
\begin{barticle}
\bauthor{\bsnm{Piepho}, \binits{H.-P.}},
\bauthor{\bsnm{Richter}, \binits{C.}},
\bauthor{\bsnm{Spilke}, \binits{J.}},
\bauthor{\bsnm{Hartung}, \binits{K.}},
\bauthor{\bsnm{Kunick}, \binits{A.}},
\bauthor{\bsnm{Th{\"o}le}, \binits{H.}}:
\batitle{Statistical aspects of on-farm experimentation}.
\bjtitle{Crop and Pasture Science}
\bvolume{62},
\bfpage{721}--\blpage{735}
(\byear{2011})
\doiurl{10.1071/CP11175}
\end{barticle}
\endbibitem

\bibitem[\protect\citeauthoryear{Kim et~al.}{2019}]{kim2019unmanned}
\begin{barticle}
\bauthor{\bsnm{Kim}, \binits{J.}},
\bauthor{\bsnm{Kim}, \binits{S.}},
\bauthor{\bsnm{Ju}, \binits{C.}},
\bauthor{\bsnm{Son}, \binits{H.I.}}:
\batitle{Unmanned aerial vehicles in agriculture: A review of perspective of
  platform, control, and applications}.
\bjtitle{Ieee Access}
\bvolume{7},
\bfpage{105100}--\blpage{105115}
(\byear{2019})
\end{barticle}
\endbibitem

\bibitem[\protect\citeauthoryear{Weiss et~al.}{2020}]{weiss2020remote}
\begin{barticle}
\bauthor{\bsnm{Weiss}, \binits{M.}},
\bauthor{\bsnm{Jacob}, \binits{F.}},
\bauthor{\bsnm{Duveiller}, \binits{G.}}:
\batitle{Remote sensing for agricultural applications: A meta-review}.
\bjtitle{Remote Sensing of EEnvironment}
\bvolume{236},
\bfpage{111402}
(\byear{2020})
\end{barticle}
\endbibitem

\bibitem[\protect\citeauthoryear{Benami et~al.}{2021}]{benami2021uniting}
\begin{barticle}
\bauthor{\bsnm{Benami}, \binits{E.}},
\bauthor{\bsnm{Jin}, \binits{Z.}},
\bauthor{\bsnm{Carter}, \binits{M.R.}},
\bauthor{\bsnm{Ghosh}, \binits{A.}},
\bauthor{\bsnm{Hijmans}, \binits{R.J.}},
\bauthor{\bsnm{Hobbs}, \binits{A.}},
\bauthor{\bsnm{Kenduiywo}, \binits{B.}},
\bauthor{\bsnm{Lobell}, \binits{D.B.}}:
\batitle{Uniting remote sensing, crop modelling and economics for agricultural
  risk management}.
\bjtitle{Nature Reviews Earth \& Environment}
\bvolume{2}(\bissue{2}),
\bfpage{140}--\blpage{159}
(\byear{2021})
\end{barticle}
\endbibitem

\bibitem[\protect\citeauthoryear{Wulder et~al.}{2022}]{wulder2022fifty}
\begin{barticle}
\bauthor{\bsnm{Wulder}, \binits{M.A.}},
\bauthor{\bsnm{Roy}, \binits{D.P.}},
\bauthor{\bsnm{Radeloff}, \binits{V.C.}},
\bauthor{\bsnm{Loveland}, \binits{T.R.}},
\bauthor{\bsnm{Anderson}, \binits{M.C.}},
\bauthor{\bsnm{Johnson}, \binits{D.M.}},
\bauthor{\bsnm{Healey}, \binits{S.}},
\bauthor{\bsnm{Zhu}, \binits{Z.}},
\bauthor{\bsnm{Scambos}, \binits{T.A.}},
\bauthor{\bsnm{Pahlevan}, \binits{N.}}, \betal:
\batitle{Fifty years of landsat science and impacts}.
\bjtitle{Remote Sensing of Environment}
\bvolume{280},
\bfpage{113195}
(\byear{2022})
\end{barticle}
\endbibitem

\bibitem[\protect\citeauthoryear{Argento et~al.}{2021}]{argento2021}
\begin{barticle}
\bauthor{\bsnm{Argento}, \binits{F.}},
\bauthor{\bsnm{Anken}, \binits{T.}},
\bauthor{\bsnm{Abt}, \binits{F.}},
\bauthor{\bsnm{Vogelsanger}, \binits{E.}},
\bauthor{\bsnm{Walter}, \binits{A.}},
\bauthor{\bsnm{Liebisch}, \binits{F.}}:
\batitle{Site-specific nitrogen management in winter wheat supported by
  low-altitude remote sensing and soil data}.
\bjtitle{Precision Agriculture}
\bvolume{22}(\bissue{2}),
\bfpage{364}--\blpage{386}
(\byear{2021})
\doiurl{10.1007/s11119-020-09733-3}
\end{barticle}
\endbibitem

\bibitem[\protect\citeauthoryear{Roth et~al.}{2018}]{Roth2018a}
\begin{botherref}
\oauthor{\bsnm{Roth}, \binits{L.}},
\oauthor{\bsnm{Hund}, \binits{A.}},
\oauthor{\bsnm{Aasen}, \binits{H.}}:
{{PhenoFly Planning Tool}}: Flight planning for high-resolution optical remote
  sensing with unmanned areal systems.
Plant Methods
\textbf{14}(116)
(2018)
\doiurl{10.1186/s13007-018-0376-6}
\end{botherref}
\endbibitem

\bibitem[\protect\citeauthoryear{Tummers et~al.}{2019}]{Tummers2019}
\begin{barticle}
\bauthor{\bsnm{Tummers}, \binits{J.}},
\bauthor{\bsnm{Kassahun}, \binits{A.}},
\bauthor{\bsnm{Tekinerdogan}, \binits{B.}}:
\batitle{Obstacles and features of {{Farm Management Information Systems}}:
  {{A}} systematic literature review}.
\bjtitle{Computers and Electronics in Agriculture}
\bvolume{157},
\bfpage{189}--\blpage{204}
(\byear{2019})
\doiurl{10.1016/j.compag.2018.12.044}
\end{barticle}
\endbibitem

\end{thebibliography}

\end{document}